\documentclass[aps,10pt,twocolumn,floats,floatfix,prl,reprint,longbibliography]{revtex4-2}
\usepackage[T1]{fontenc}
\usepackage[utf8]{inputenc}
\usepackage{amsmath, amssymb, amsthm}
\usepackage{graphicx}
\usepackage{orcidlink}
\usepackage{tikz,pgf}
\usepackage{pgfplots}
\pgfplotsset{compat=1.18}
\usepackage{subfigure}
\usepackage{color}
\usepackage{xcolor}
\usepackage{bm,bbm}
\usepackage{enumerate}
\usepackage{braket}
\usepackage{hyperref}
\hypersetup{
    colorlinks=true,
    linkcolor=blue,
    citecolor=blue,
    urlcolor=blue,
    filecolor=blue
}

\usepackage{mathptmx}

\begin{document}

\title{Semiclassical Approach to Quantum Fisher Information}

\author{Mahdi RouhbakhshNabati\,\orcidlink{0009-0006-3288-3372}}
\email[Contact author: ]{mahdi@rouhbakhsh.net}
\affiliation{Institut für Theoretische Physik, Eberhard-Karls-Universität Tübingen, 72076 Tübingen, Germany}
\author{Daniel Braun\,\orcidlink{0000-0001-8598-2039}}
\email[Contact author: ]{daniel.braun@uni-tuebingen.de}
\affiliation{Institut für Theoretische Physik, Eberhard-Karls-Universität Tübingen, 72076 Tübingen, Germany}
\author{Henning Schomerus\,\orcidlink{0000-0002-7959-0992}}
\email[Contact author: ]{h.schomerus@lancaster.ac.uk}
\affiliation{Department of Physics, Lancaster University, Lancaster, LA1 4YB, United Kingdom}
\date{\today}

\begin{abstract}
Quantum sensors driven into the quantum chaotic regime can have dramatically enhanced sensitivity, which, however, depends intricately on the details of the underlying classical phase space. Here, we develop an accurate semiclassical approach that provides direct and efficient access to the phase-space-resolved quantum Fisher information (QFI), the central quantity that quantifies the ultimate achievable sensitivity. This approximation reveals, in very concrete terms, that the QFI is large whenever a specific dynamical quantity tied to the sensing parameter displays a large variance over the course of the corresponding classical time evolution. Applied to a paradigmatic system of quantum chaos, the kicked top, we show that the semiclassical description is accurate already for modest quantum numbers, i.e., deep in the quantum regime, and it extends seamlessly to very high quantum numbers that are beyond the reach of other methods.
\end{abstract}

\maketitle

\emph{Introduction---}
Quantum Fisher information (QFI)
plays a crucial role in quantum metrology \cite{Helstrom1969, HOLEVO1973337, Gibilisco2007, PhysRevLett.72.3439,braunstein_generalized_1996, article,fiderer_quantum-chaotic_2019}. Its inverse sets the quantum Cramér-Rao bound, the ultimate achievable lower bound on the variance of an unbiased estimator, optimized over all possible positive-operator-valued-measure measurements and data analysis schemes with unbiased estimators. The higher the QFI with respect to a certain parameter on which the quantum state depends, the better that parameter can be estimated, making QFI essential for the development of quantum sensors that outperform classical limitations. 
In a variety of multipartite quantum systems, superlinear scaling of the QFI with the number of constituents can be directly linked to the presence of quantum entanglement \cite{PhysRevLett.102.100401,Giovannetti2011}, but quantum-enhanced metrology schemes that do not rely on entanglement exist, notably for parameters that characterize the interactions between subsystems \cite{Boixo08,Boixo08.2,fraisse_coherent_2015,RevModPhys.90.035006}. In \cite{Fiderer2018,fiderer_quantum-chaotic_2019} it was shown that imprinting the parameter via quantum-chaotic dynamics can lead to largely enhanced sensitivity. Understanding the mechanisms for the enhancement, then, poses a significant challenge, as it depends on the intricate connection between classical phase space structures and the phase-coherent dynamics of the quantum system. This state of affairs is further acerbated by the fact that, in general, one is interested in the regime of large quantum numbers, as QFI in many cases increases with the relevant quantum numbers (such as the photon number in interferometric schemes or total angular momentum in magnetometry). Large quantum numbers motivate the use of semiclassical techniques, in particular if they place the system out of reach of direct quantum-mechanical approaches.
These connections position a semiclassical theory of QFI as a challenging but vital step in understanding and exploring the mechanisms by which nonlinear dynamics can enhance the sensitivity of quantum sensors.

Here, we introduce such a semiclassical approach to obtain the QFI and demonstrate that it is highly accurate and efficient.  
The approach ties the QFI in very concrete and simple terms to the variance of a classical dynamical observable associated with the estimation parameter and precisely determines the procedure by which this quantity is to be evaluated. This offers a computationally efficient and accurate alternative to conventional quantum calculations for nonintegrable quantum systems, specifically when numerical calculations become impractical due to a large Hilbert space dimension.

\emph{Setting and main result---} 
The QFI of a parameter-dependent pure quantum state $\ket{\psi}=\ket{\psi(\beta)}$ is given by \cite{PhysRevLett.72.3439,braunstein_generalized_1996}
\begin{align}
I_\beta = 4 \, (\, \braket{\partial_{\beta}\psi|\partial_{\beta}\psi}-|\braket{\psi|\partial_{\beta}\psi}|^2\, )\,,
\label{eq:QFI0}
\end{align}
where $\beta$ is the estimation parameter, and $\partial_\beta\equiv \partial/\partial\beta$ denotes the partial derivative. We are interested in the setting where $\ket{\psi(\beta)}= \ket{\psi(\beta,t)}=U_\beta(t)\ket{\psi_0}$ arises from the evolution out of a given initial state $\ket{\psi(0)}=\ket{\psi_0}$, where $\beta$ enters the time-evolution operator $U_\beta(t)$. For a given dynamical system, $I_\beta(\ket{\psi_0},t)$ can then be interpreted as a time-dependent quantity that depends on the initial conditions. The suitability of the system for quantum sensing can be analyzed in detail by varying the initial state and duration of the time evolution. Furthermore, the relation to classical nonlinear dynamics can be investigated by selecting initial states $\ket{\psi_0}=\ket{\mathbf z_0}$ that correspond to a classical phase space location $\mathbf{z}_0$, as specified further below. For a system with nonlinear classical dynamics, this initial state will quickly become distorted and develop a complex interference pattern, which we can analyze by studying $I_\beta(\mathbf{z}_0,t)\equiv I_\beta(\ket{\mathbf{z}_0},t)$.

To obtain the corresponding semiclassical description, we therefore have to find suitable representations for both the initial conditions encoded in $\ket{\psi_0}$ and the phase-coherent dynamics in $U_\beta(t)$.
This can be achieved within a uniform approximation that deals with both parts of this problem at the same time until they decouple again in the final result. To set this up, we exploit the link to another well-known quantity in quantum chaos capturing the sensitivity of the dynamics with respect to a parameter,
the Loschmidt echo \cite{gorin_dynamics_2006}:
\begin{align}
\chi&=\braket{ \psi_0 | U^\dagger_{\beta-\varepsilon}(t)U_{\beta+\varepsilon}(t)|\psi_0 } .
\end{align}
It reflects how well the system can recover its initial state if, after time $t$, the dynamics with parameter $\beta+\varepsilon$ is reversed and the state propagated back with a slightly changed parameter $\beta-\varepsilon$. In terms of the Loschmidt echo, the QFI can be exactly rewritten as $I=\chi'^2-\chi''$, where the derivatives are with respect to the parameter $\varepsilon$, which is finally set to 0 \cite{Fiderer2018}.
Now, the \textit{composed propagator} $U^\dagger_{\beta-\varepsilon}(t)U_{\beta+\varepsilon}(t)$ can be expressed as a path integral and approximated semiclassically using the Van Vleck-Gutzwiller (VVG) propagator \cite{Gutzwiller1990}, which can furthermore be evaluated directly in the path-integral representation of the initial state \cite{Note1}.
This combines the two mentioned tasks into one. Expanded in the relevant orders in $\varepsilon$, we then obtain a uniform semiclassical approximation of the echo amplitude, which for $f$ degrees of freedom, is given as follows \footnote{See supplemental material available with this submission for further details of the derivation of the semiclassical QFI and its application to the QKT, the quantum kicked rotor \cite{IZRAILEV1990299,tworzydlo2003dynamical}, and H{\'e}non-Heiles system \cite{brack2018semiclassical}.}
\begin{align}
\chi\approx&\frac{1}{(\pi\hbar)^f}\int d^{2f}\mathbf{z}
\exp\left[-\frac{(\mathbf{z}-\mathbf{z}_0)^2}{\hbar}+ \frac{i}{\hbar}2\varepsilon \frac{\partial S}{\partial \beta}\right].
\label{eq:finalformchi}
\end{align}
Here, the integral is over the initial phase space points $\mathbf{z}$, whose localization saturates the Heisenberg uncertainty limit, while $S$ is the classical action of the trajectory from the initial to the final time of the dynamics, which is the same action as used in the path integral representation of the propagator $U_\beta(t)$ (not the composed propagator). 

Using the abovementioned link to the QFI, we then find it semiclassically expressed as
\begin{align}
I_{\mathrm{sc}}(\mathbf{z}_0,t)&=\frac{4}{\hbar^2}\mathrm{var}\left(\frac{\partial S}{\partial \beta}\right),
\label{eq:mainVVG1}
\end{align}
which is the main general result of our work.
Thus, semiclassically, the QFI is expressed in terms of the variance of a particular classical dynamical quantity, the action derivative $\frac{\partial S}{\partial \beta}$, which is to be evaluated from classical initial conditions that optimally align with the chosen initial state (for our initial wave packets $\ket{\mathbf{z}_0}$, a Gaussian distribution in phase space that saturates the Heisenberg uncertainty principle;
other semiclassical approaches, such as those based on the Husimi function, deviate from these optimal conditions).

Even though this involves classical data, the final result is an approximation of the quantum Fisher information, rather than a classical Fisher information, which is always linked to a specific measurement. It is ``quantum'' in the sense that it contains quantum interference terms present in the propagator, even though only paths close to the classical path contribute to the semiclassical VVG propagator. The approximation becomes exact for a classically linear dynamics of the state, which retains its Gaussian nature. In this special case, our approach delivers the exact quantum result $I_{\mathrm{sc}}(\mathbf{z}_0,t)=I(\mathbf{z}_0,t)$.

On the other hand, because our result involves only classical data, it can be evaluated efficiently. Along with the observation that the action derivatives are typically simple quantities, we posit that the result [Eq.~\eqref{eq:mainVVG1}] can be efficiently applied to a wide range of systems. 
To further assert the validity and efficiency of our approach in a nonlinear setting, we now apply it to a paradigmatic model system.

\emph{Application to a paradigm of quantum chaos---}
To demonstrate the power of our method, we apply it to a widely studied model of quantum chaos with immediate applications to quantum sensing, the quantum kicked top (QKT) \cite{haake1987classical,Haake1991,chaudhury_quantum_2009,Fiderer2018}. It is composed of coupled spins with a conserved total spin magnitude $\hbar\sqrt{J(J+1)}$, where the total angular momentum quantum number $J$ can be an integer or half integer. 
We work in units in which the collective spin operators $J_x, J_y, J_z$ follow the commutation relations $[J_l,J_m]=i \epsilon_{lmn}J_n$ for $l,m,n \in \{x,y,z\}$, which amounts to setting $\hbar\equiv 1$. The dynamics is governed by periodic kicks, where the unitary propagator over one period $T$ is given by
\begin{align}
U_{\beta}(T) = \exp\left(-i k \frac{J_z^2}{2J+1}\right)\exp{(-i\beta J_y)},
\label{eq:one_step_unitary}
\end{align}
while the subsequent stroboscopic motion follows from $U_\beta(t\,T)=U^t_{\beta}(T)$, with integer $t\geq 1$. The propagator $U_{\beta}(T)$ consists of two sequential operations: a free precession about the $y$-axis by an angle $\beta$, followed by a nonlinear torsion (kick) about the $z$-axis, which is controlled by the parameter $k$. Quantum mechanically, these dynamics unfold in a Hilbert space of dimension $2J+1$. In the classical limit $J\to\infty$, $\{J_x/(J+1/2), J_y/(J+1/2), J_z/(J+1/2)\}\to \{x,y,z\}$ become the Cartesian coordinates of a unit vector, for which the torsion can be interpreted as a position-dependent rotation by an angle $k\,z$. This classical motion can be conveniently studied in a 2D phase space $(\phi,z)$, where $\phi=\arctan{(y/x)}$ is the azimuthal angle of the unit vector. For large enough $k$, the nonlinear kicks drive the system into the chaotic regime. 

The very same classical motion is also embedded into the semiclassical approximation of the propagator in Eq.~\eqref{eq:one_step_unitary}, which identifies the action $S$ with the generator of the classical dynamics \cite{Braun1996,Note1}. Given our rescaling of the angular momentum vector to a unit vector, the effective reduced Planck constant entering the semiclassical description is given by $\hbar_\mathrm{eff}=1/(J+1/2)$ \footnote{We recall that we have set the physical reduced Planck constant $\hbar=1$ in the angular momentum commutation relations. In addition to tracking its definition through rescaling of the angular momentum vector, the stated effective value also agrees with the Weyl rule, according to which the Hilbert space dimension $2J+1$ equals the phase space volume $2\times 2\pi$ divided by the size $h_\mathrm{eff}=2\pi/(J+1/2)$ of a Planck cell.}.

\begin{figure*}[t] % Use figure* for spanning both columns
    \centering
    \subfigure{%
         \includegraphics[width=0.31\textwidth]{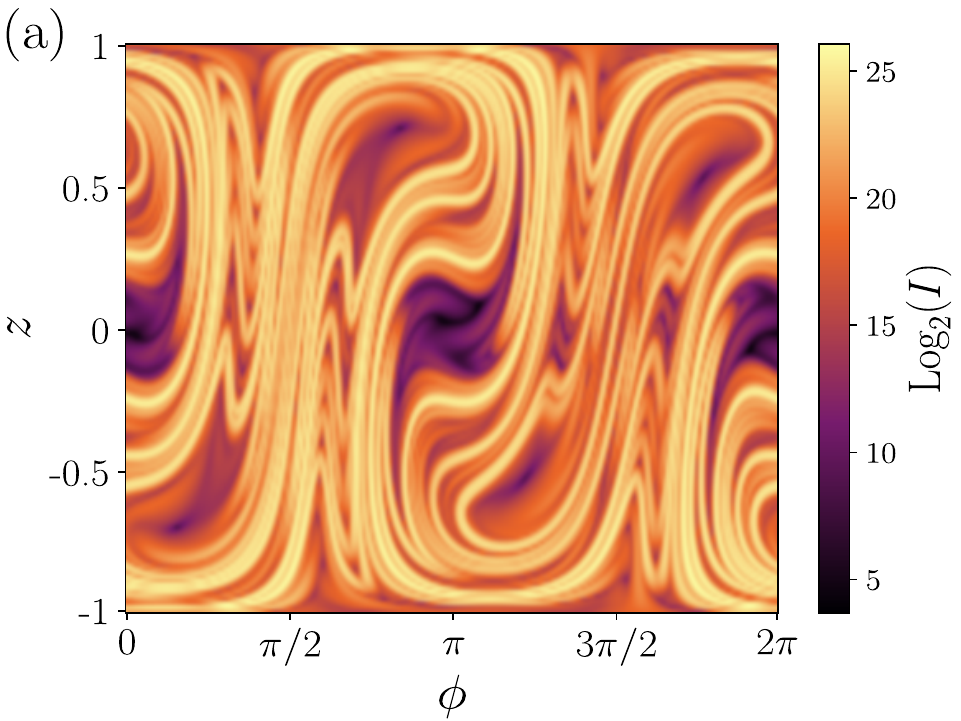}%
        \label{fig:first}%
    }\hfill
    \subfigure{%
        \includegraphics[width=0.31\textwidth]{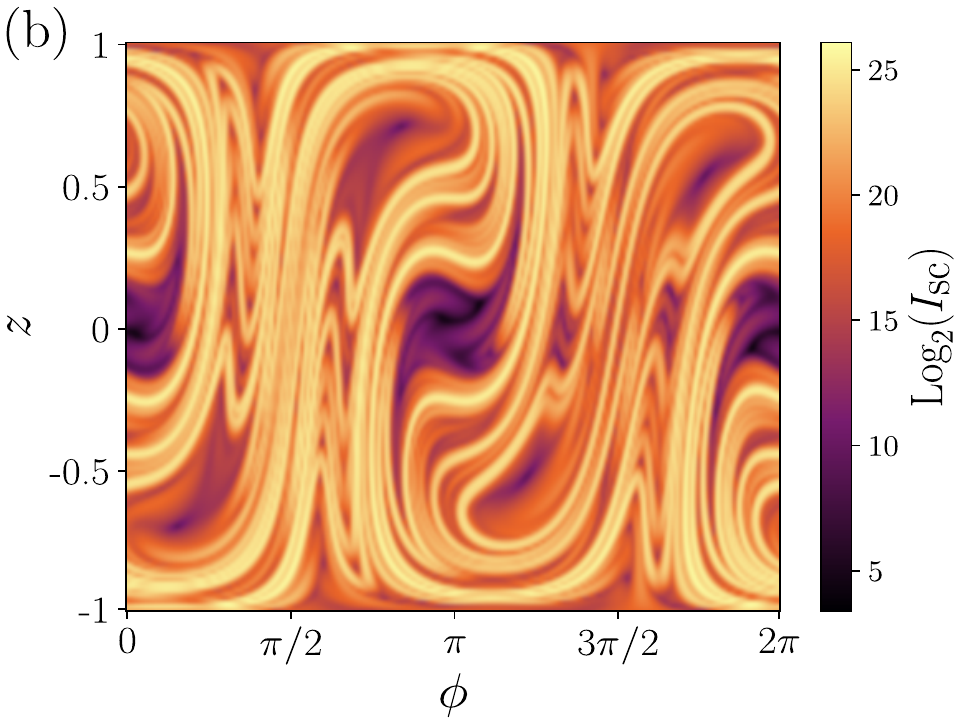}%
        \label{fig:second}%
    }\hfill
    \subfigure{%
        \includegraphics[width=0.31\textwidth]{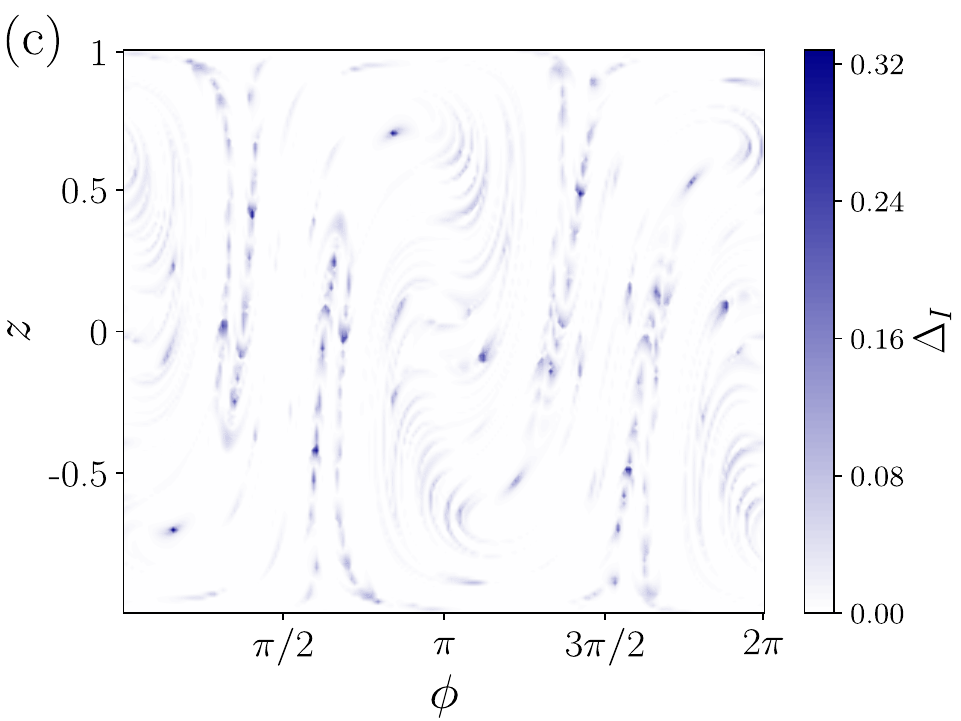}%
        \label{fig:third}%
    }
    \caption{\emph{Comparison of the semiclassical QFI, Eq.~\eqref{eq:mainVVG1}, with the exact QFI for the quantum kicked top as a function of the position of the initial $\mathrm{SU}(2)$ coherent state.} The initial coherent state is parametrized in terms of the phase space variables $\phi\in(0,2\pi)$ and $z\in(-1,1)$, which are discretized using 220 and 150 equidistant points, respectively, in the $\phi$ and $z$ directions. The calculations are performed for $J=4096$, $t=8$, $\beta=1.5$, and $k=3$. The portraits of the exact QFI in (a) and the semiclassical result in (b) are in excellent agreement.
    Both quantities are plotted on a logarithmic scale to cover their wide range of values and verify agreement into regions where they are both small.
    (c) Quantifies the difference between the two results in terms of the relative deviation $\Delta_I=\left|{\bar I}-{\bar I}_\mathrm{sc}\right|/({\bar I} + {\bar I}_\mathrm{sc})$.}
    \label{fig:phase_space_comp}
\end{figure*}

For the application to the semiclassical QFI [Eq.~\eqref{eq:mainVVG1}], we require knowledge of the action derivative over only a single time step $T\equiv 1$ \cite{Note1, Braun1996},
\begin{align}
\frac{\partial S^{(1)}(z_t,z_{t+1})}{\partial \beta} = -\sqrt{1-z_t^2}\sin{\phi_t}=-y_t
\end{align}
from which the complete action derivative follows as
\begin{align}
\frac{\partial S(t)}{\partial \beta} = \sum_{t'=0}^{t-1} \frac{\partial S^{(1)}(z_{t'},z_{t'+1})}{\partial \beta} = -\sum_{t'=0}^{t-1} y_{t'}.
\label{eq:action_summation}
\end{align}
Therefore, the semiclassical QFI amounts to calculating
\begin{equation}
    I_\mathrm{sc}(\mathbf{z}_0,t)=(2J+1)^2\;\mathrm{var}\,\left(
    \sum_{t'=0}^{t-1}y_{t'}\right),
\end{equation}
where the variance is evaluated with the stated classical initial conditions.

This fixes the dynamical content of our semiclassical description, so that we can now turn to the initial conditions required to calculate the variance in Eq.~\eqref{eq:mainVVG1}.
Quantum mechanically, we specify the initial states $\ket{\mathbf{z}_{0}}=\ket{J,\theta,\phi}$ with $\theta=\arccos{(z)}$ as $\mathrm{SU}(2)$ spin-coherent states \cite{Arecchi72,Agarwal86,Dowling94,perelomov_generalized_1977,Haake1991},
\begin{align}
\label{eq:spin_coherent_state}
&\ket{J,\theta,\phi}\\
&=\sum_{m=-J}^{J}\sqrt{
\begin{pmatrix}
2J\\
J-m
\end{pmatrix}
}[e^{i\phi}\sin{(\frac{\theta}{2})}]^{J-m}\cos{(\frac{\theta}{2})^{J+m}}\ket{J,m}
,
\nonumber
\end{align}
which are the most localized quantum states in phase space that saturate the Heisenberg uncertainty relation. Following our derivation of Eqs.~\eqref{eq:finalformchi} and \eqref{eq:mainVVG1}, their semiclassical approximation then translates into a corresponding Gaussian distribution in phase space, again with the effective reduced Planck constant $\hbar_\mathrm{eff}=1/(J+1/2)$.

\emph{Parameter estimation across phase space---}
We now present numerical results that compare the exact QFI $I(\mathbf{z}_0,t)$ with the semiclassical approximation $I_{\mathrm{sc}}(\mathbf{z}_0,t)$.
For the exact QFI, we evolve the initial spin-coherent state in Eq.~\eqref{eq:spin_coherent_state} with the unitary time evolution operator in Eq.~\eqref{eq:one_step_unitary} and employ an efficient recursive method involving the parametric derivative $\partial_{\beta} U_\beta$ \cite{Note1}. 
For the semiclassical method, we obtain the variance by evaluating points of linear spacing $\sigma/r$, weighted with a symmetric Gaussian distribution of width $\sigma=(2J+1)^{-1/2}$ centered at $\mathbf{z}_0$, where the parameter $r$ serves as the phase space resolution parameter. At this stage, we set $r=50$, but revisit the role of this parameter further below (see also \cite{Note1}).

\begin{figure*}[t] % Use figure* for spanning both columns
    \centering
    \hspace{-0.16cm} %
    \subfigure{
        \includegraphics[width=0.25\textwidth]{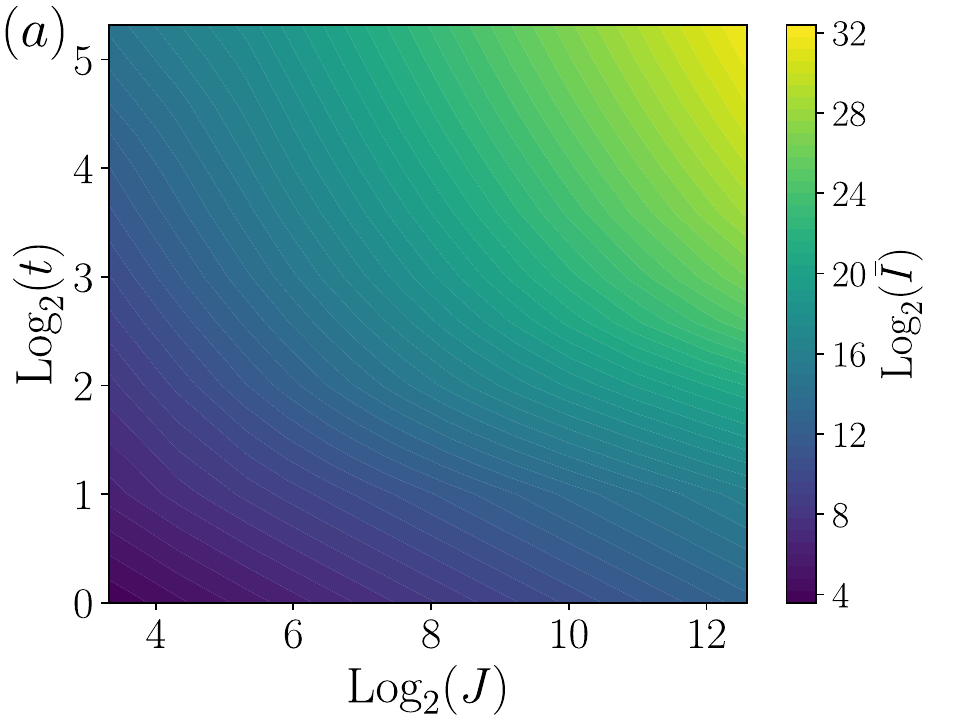}%
    }%
    \subfigure{%
        \includegraphics[width=0.25\textwidth]{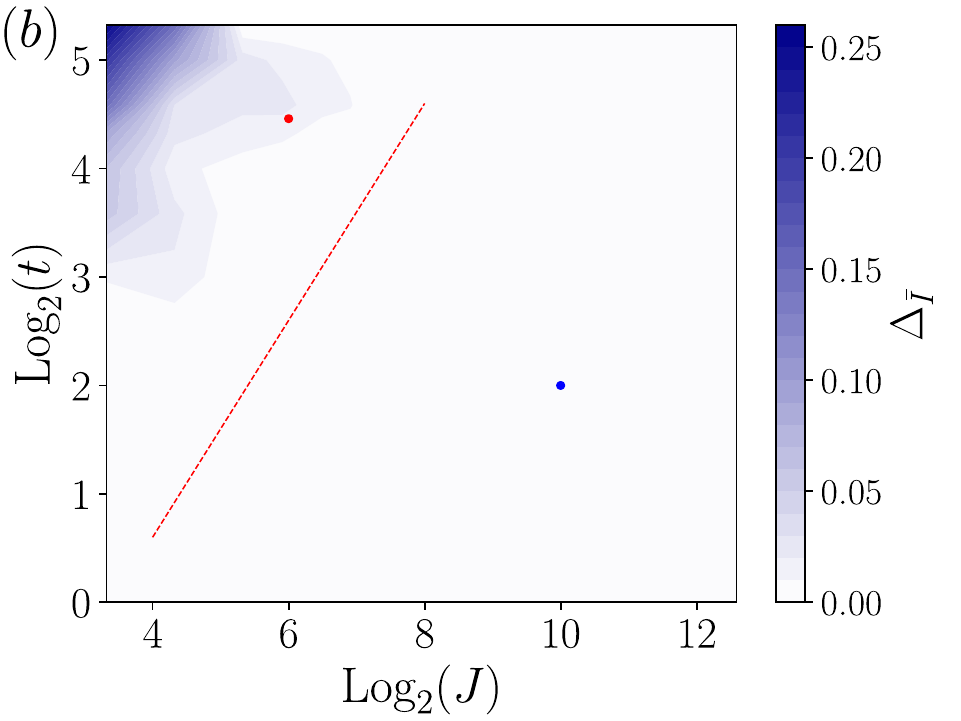}}%
    \subfigure{%
        \includegraphics[width=0.25\textwidth]{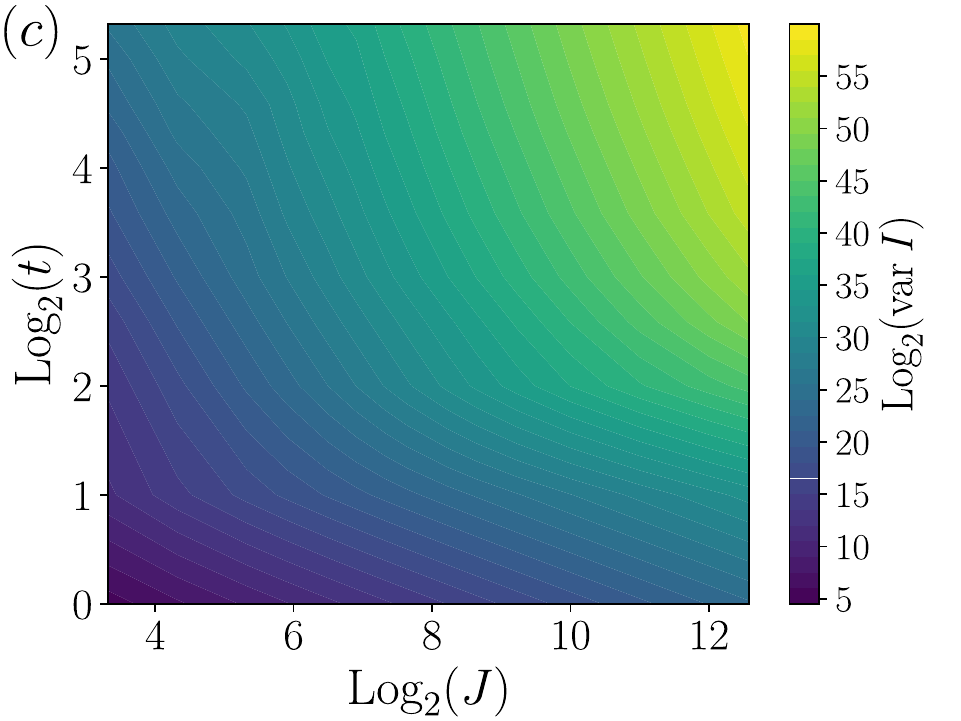}
    }%
    \subfigure{%
        \includegraphics[width=0.25\textwidth]{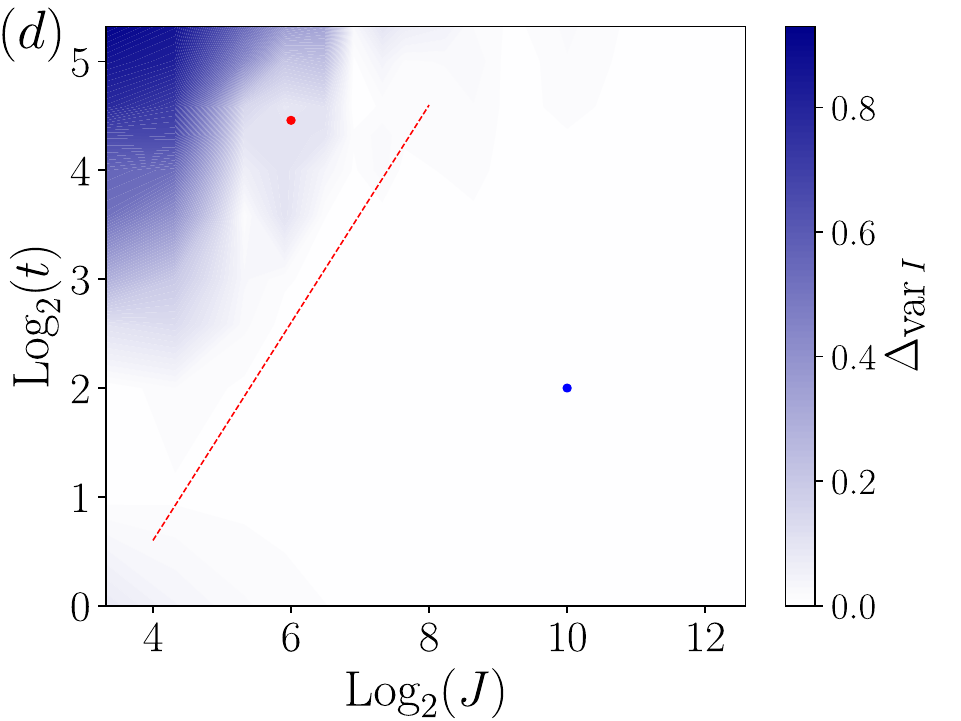}}

    \caption{\emph{Comparison of the exact and semiclassical results for the QFI of the QKT in terms of the phase space average $\bar I$ and the phase space variance $\mathrm{var}\, I$, as a function of $J$ and $t$ for fixed $k=4$.}
    All quantities are plotted on a logarithmic scale to cover a wide range of values and parameters. (a),(c) Heat maps of $\bar I$ and $\mathrm{var}\,I$ from the exact QFI. (b),(d) Deviation of the corresponding results of the semiclassical approach from the exact values. According to panels (a) and (c), the average QFI and its fluctuations in phase space both increase with $t$ and $J$. The white areas in panels (b) and (d) demonstrate excellent agreement of the semiclassical method in the semiclassical regime of large $J$. 
   The red dashed lines indicate scaling with the Heisenberg time $t_{H}\propto J$.
   }
    \label{fig:ave_var_comp_fixed_kz}
\end{figure*}

Figure~\ref{fig:phase_space_comp} shows heat maps of the exact and semiclassical QFI as a function of the location of the initial state in phase space, calculated for $J=4096$, $t=8$, $\beta=1.5$, and $k=3$. The right plot shows the relative numerical difference,
\begin{equation}
    \Delta_I=\left|I-I_{\mathrm{sc}}\right|/(I+I_{\mathrm{sc}}),
    \label{eq:comparison}
\end{equation}
between these two methods across the phase space. The visible agreement is excellent, while the numerical difference shows a close to perfect match for the semiclassical method except for a few locations, which correlate to initial positions that yield a very small or rapidly varying exact QFI.

To assess the overall agreement over a larger range of parameters, we introduce two global statistical measures, the average $\bar I\equiv\overline{I(\mathbf{z}_0,t)}$ and the variance $\mathrm{var}\,I\equiv\mathrm{var}\,I(\mathbf{z}_0,t)$ of the QFI over phase space, along with their relative differences $\Delta_{\bar I}=\left|{\bar I}-{\bar I}_\mathrm{sc}\right|/({\bar I} + {\bar I}_\mathrm{sc})$ and 
$\Delta_{\mathrm{var}\, I}=\left|{\mathrm{var}\, I}-{\mathrm{var}\, I}_\mathrm{sc}\right|/({\mathrm{var}\, I}+{\mathrm{var}\, I}_\mathrm{sc})$. 
For the parameters of the phase space portraits discussed, the relative differences, $\Delta_{\bar I}=0.000327$ and $\Delta_{\mathrm{var}\, I}=0.000425$, are very small.

Figure \ref{fig:ave_var_comp_fixed_kz} uses these global measures to quantify the accuracy of the semiclassical approach over a wider range of the angular momentum parameter $J$ and time $t$, with the linear rotation parameter $\beta=1.5$ and nonlinear torsion $k=4$ kept fixed at a representative value of a mixed phase space.
Panels (a) and (c) show triple-logarithmic heat maps of $\bar I$ and $\mathrm{var}\,I$, while panels (b) and (d) show the corresponding deviations.
The angular momentum parameter $J$ covers the range from $10$ up to $6144$, limited by the Hilbert space dimension $2J+1$ in the exact calculation, while time $t$ covers the range from $1$ to $40$. The logarithmic scale is chosen, since $\bar I$ and $\mathrm{var}\, I$ vary over a very wide range of values but exhibit a smooth and systematic dependence on $J$ and $t$.

The figure demonstrates that the semiclassical approximation becomes increasingly precise for larger values of $J$, hence, as one approaches the classical limit. 
Furthermore, we observe that in this limit, the time horizon of the semiclassical approximation scales with the Heisenberg time $t_H=2J+1$ \cite{Haake1991},
as indicated by the red line $\propto J$. We observed similar behavior in other models of quantum chaos \cite{Note1}.

\emph{Efficiency of the method---}
In the numerical implementation of the method, the semiclassical method is highly efficient and is mainly limited by the phase space resolution that is required to obtain a sufficiently accurate estimate of the variance in Eq.~\eqref{eq:mainVVG1}. In particular, we observe that the semiclassical approach significantly outperforms the exact quantum-mechanical calculations when the phase space is mixed; hence, it displays a highly distinctive parameter-dependent structure.

\begin{figure}[b] % Use figure* for spanning both columns
    \centering
    \hspace{-0.15cm} %
    \subfigure{
        \includegraphics[width=0.23\textwidth]{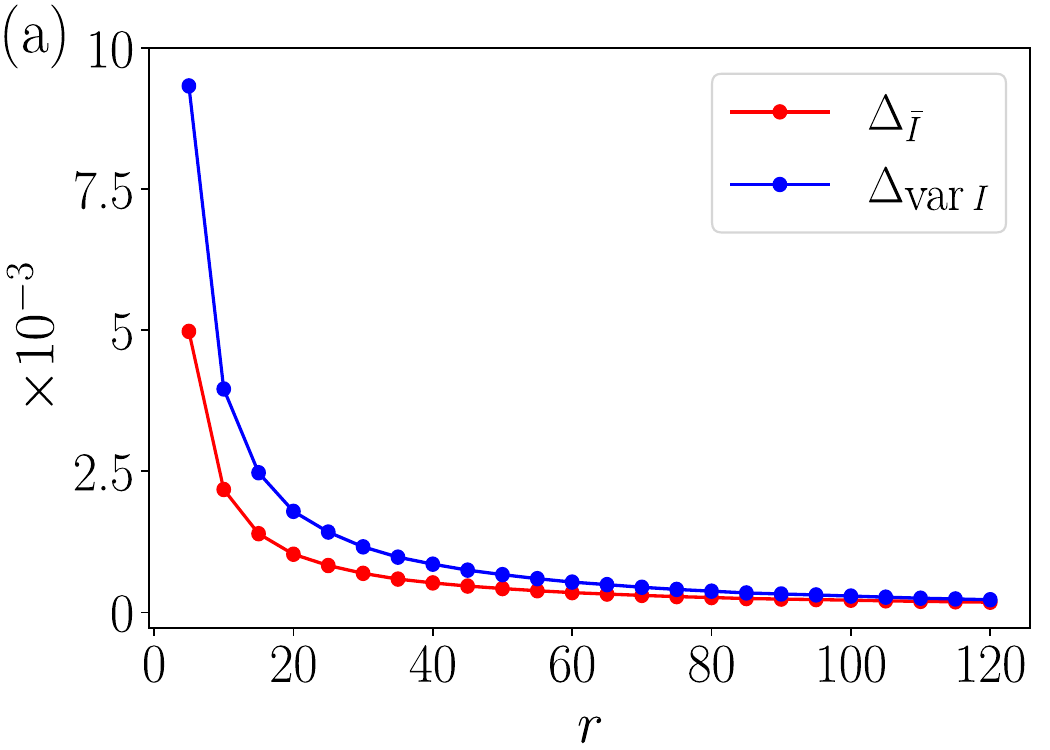}%
    }\hfill
    \subfigure{%
        \includegraphics[width=0.23\textwidth]{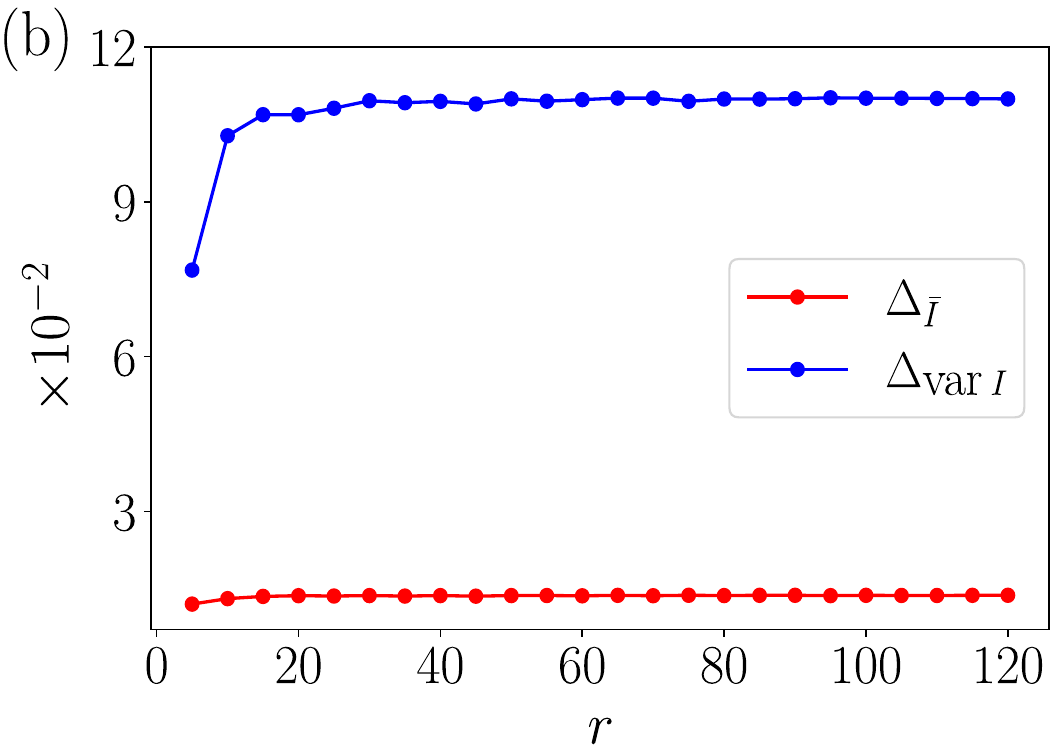}%
    }
    \caption{Convergence of the averaged semiclassical QFI and its variance as a function of the phase space resolution parameter, $r$, for two different conditions in Fig.~\ref{fig:ave_var_comp_fixed_kz} (where $k=4$). In (a), $J=1024$ and $t=4$, as indicated by the blue dot in Figs.~\ref{fig:ave_var_comp_fixed_kz}(b) and \ref{fig:ave_var_comp_fixed_kz}(d). 
    This corresponds to the regime where the semiclassical description is highly accurate. Panel (b) contrasts this with the case $J=64$ and a larger time $t=22$, indicated by the red dot in Figs.~\ref{fig:ave_var_comp_fixed_kz}(b) and \ref{fig:ave_var_comp_fixed_kz}(d), where the semiclassical method provides a less accurate, but still reasonable, estimate.}
    \label{fig:increasing_r}
\end{figure}

We can quantify the comparative efficiency of the exact and approximate methods in terms of the scaling of the computational effort with $J$ and $t$.
The computational effort of the exact QFI is dominated by matrix multiplications in the calculation of the propagator, regardless of whether we consider a single phase space point or the entire phase space (partitioned in Planck cells). Without specialized treatment, the matrix multiplications scale as $J^3$ (specialized treatments achieve a scaling below the third power, such as Strassen's algorithm, which scales as $\sim J^{2.807}$ \cite{strassen_gaussian_1969}). In contrast, the semiclassical approach scales with $J^1$ if one considers the complete phase space and $J^0$ if one considers a given fixed point.
Important, however, is the factor of this scaling, which depends on the complexity of the phase space. This can be assessed by varying the phase space resolution parameter $r$.
Our results, with an example shown in Fig.~\ref{fig:increasing_r}, support the accuracy of the semiclassical method in the mixed phase space regime.

A competing picture arises when one considers the computational effort as a function of time. The matrix multiplications in the quantum case can be organized hierarchically to obtain logarithmic scaling $\sim \log_2(t)$ \cite{Note1}. Semiclassically, the bare numerical effort scales as $t$.
In the strongly chaotic regime, this is further exacerbated by the strong sensitivity of the trajectories to the initial conditions. Practically, at large times the numerical implementation of the semiclassical method then captures only the universal component of the ensuing fluctuations. However, our results in Fig.~\ref{fig:increasing_r} (b) indicate that these still show reasonable agreement with the exact result, which suggests that the universal contribution is dominant, as also exhibited in the exact QFI.

Further efficiencies of the semiclassical treatment could be gained by propagating the linearized motion along with the trajectory, and hence, gaining insight about the contributions from the vicinity of a given point. This could furthermore be complemented with statistical and analytical insights into universal behavior that emerges at long times and small scales of the phase space. We have not explored such further improvements and insights, leaving them as topics for future research.

\emph{Conclusions---}
We introduced a semiclassical method to calculate the quantum Fisher information, which amounts to evaluating the variance of the parametric derivative of the action of classical trajectories weighted with a phase space distribution determined by the initial quantum state. Our numerical results for a paradigm of quantum chaos, the kicked top,  demonstrate that the method is highly efficient compared to the exact computation for the quantum system, and it produces reliable results up to the Heisenberg time. 
This addresses major bottlenecks in the calculation of the QFI, which, when performed exactly, requires dealing with operators in the Hilbert space, and hence is computationally demanding both in terms of memory and time.
In particular, the semiclassical method excels exactly when exact QFI computations become out of reach for large Hilbert space dimensions,
as in autonomous systems with several degrees of freedom.
Beyond these computational advantages, the approach opens a pathway to understanding metrological performance in chaotic quantum systems from a classical dynamical perspective. It also has potential applications in the design of quantum sensors and simulators, where classical modeling remains feasible, but full quantum analysis is prohibitive. Particularly worthwhile extensions of this work would concern the semiclassical description of sensing with nonclassical initial states described by partially negative quasiprobability distributions as well as the incorporation of decoherence.

\emph{Acknowledgments---}
Gefördert durch die Deutsche Forschungsgemeinschaft (DFG) --- Projektnummer 465199066. This work was funded by the Deutsche Forschungsgemeinschaft (DFG, German
Research Foundation) – Project No. 465199066.

\emph{Data availability---}
The data that support the findings of this article—including a zoomable Fig. 1 phase-space portrait and datasets for two additional systems—are openly available \cite{repository}.

%\bibliography{refs.bib}

% <for arXiv
\onecolumngrid
\clearpage

% turn numbering back on and use S1, S2, ... style
\setcounter{secnumdepth}{3}
\setcounter{section}{0}
\renewcommand\thesection{S\arabic{section}}
\renewcommand\thesubsection{S\arabic{section}.\arabic{subsection}}

\section*{Supplemental Material}
% \documentclass[aps,10pt,twocolumn,floats,floatfix,amsmath,amssymb,prr]{revtex4-2}
% \usepackage{amsmath,amssymb,amsthm}
% \usepackage{graphicx,tikz,pgf,pgfplots}
% \usepackage{subfigure}
% \usepackage{color,xcolor}
% \usepackage{bm,bbm}
% \usepackage[utf8]{inputenc}
% \usepackage{enumerate}
% \usepackage{braket}
% \usepackage{hyperref}
% \hypersetup{
%     colorlinks=true,
%     linkcolor=blue,
%     citecolor=blue,
%     urlcolor=blue,
%     filecolor=blue
% }

% \setcounter{footnote}{0}

% \usepackage{mathptmx}

% \begin{document}

% \title{Semiclassical approach to quantum Fisher Information\\ Supplemental Material}

% \author{Mahdi RouhbakhshNabati}
% \affiliation{Institut für Theoretische Physik, Eberhard-Karls-Universität Tübingen, 72076 Tübingen, Germany}
% \author{Daniel Braun}
% \affiliation{Institut für Theoretische Physik, Eberhard-Karls-Universität Tübingen, 72076 Tübingen, Germany}
% \author{Henning Schomerus}
% \affiliation{Department of Physics, Lancaster University, Lancaster, LA1 4YB, United Kingdom}
% \date{\today}

% \maketitle
% \onecolumngrid

% \vspace{1em}
% \noindent\textbf{Note:} References to “the main text” pertain to the associated \textit{Physical Review Letters} article.
% \vspace{1.5em}

\section{Loschmidt echo and quantum Fisher information}
For preparation of the derivation of the main result (4) in the main text, we demonstrate that for a quantum state $\ket{\psi(\beta)}=U_\beta (t)\ket{\psi_0}$ obtained from the evolution of an initial state $\ket{\psi_0}$, the quantum Fisher information (QFI) 
\begin{align}
I_\beta = 4 \, (\, \braket{\partial_{\beta}\psi|\partial_{\beta}\psi}-\left|\braket{\psi|\partial_{\beta}\psi}\right|^2\, )
\label{eq:QFI00APP}
\end{align}
can be obtained from Loschmidt echo amplitude
\begin{align}
\chi(\varepsilon)&=\braket{\psi_0 | U^\dagger_{\beta-\varepsilon}(t)U_{\beta+\varepsilon}(t)|\psi_0} =\braket{\psi(\beta-\varepsilon)|\psi(\beta+\varepsilon)}
\label{eq:loschmidtdef}
\end{align}
as
\begin{align}
I_\beta = {\chi'}^2-\chi'',
\label{eq:QFI1APP}
\end{align}
where we set $\varepsilon=0$ after evaluating the derivatives of $\chi$ with respect to $\varepsilon$.
\begin{proof}
From $\braket{\psi(\beta)|\psi(\beta)}=1$ we have
\begin{align}
\braket{\psi'(\beta)|\psi(\beta)}+\braket{\psi(\beta)|\psi'(\beta)}=0,\\\braket{\psi''(\beta)|\psi(\beta)}+2 \braket{\psi'(\beta)|\psi'(\beta)}+ \braket{\psi(\beta)|\psi''(\beta)}=0,
\end{align}
where primes denote derivatives with respect to $\beta$. Starting from the definition \eqref{eq:loschmidtdef} of the Loschmidt echo we obtain
\begin{align}
    \chi'=& -\braket{\psi'|\psi}+\braket{\psi|\psi'}=2 \braket{\psi|\psi'}=2i\left|\braket{\psi|\psi'}\right| ,\\
    \chi''=& \braket{\psi''|\psi}-2 \braket{\psi'|\psi'}+ \braket{\psi|\psi''}=-4 \braket{\psi'|\psi'},
\end{align}
with all derivatives of $\chi$ evaluated at $\varepsilon =0$.
Substituting these identities into the definition \eqref{eq:QFI00APP} delivers the result \eqref{eq:QFI1APP}.
\end{proof}

\section{Derivation of the main result} 
We now provide the detailed steps for the derivation of the main result (4) in the main text.
This expresses the QFI semiclassically in terms of the variance of action derivatives with respect to the estimation parameter $\beta$, where the sampling is over trajectories whose phase space distribution is determined by the initial quantum state. We first present the derivation for the setting of our main application, where this phase space distribution represents a minimal uncertainty wave packet, and then describe the generalization to other initial states.

For this derivation, we denote the center of the initial wave packet as $\mathbf{z}_0=(\mathbf{x}_0,\mathbf{p}_0)$, where $\mathbf{x}_0$ and $\mathbf{p}_0$ are two vectors of dimension $f$. 
The Loschmidt echo amplitude is then expressed as
\begin{align}
\chi&=\braket{ \mathbf z_0 | U^\dagger_{\beta-\varepsilon}U_{\beta+\varepsilon} | \mathbf z_0 }.
\end{align}
By inserting two resolutions of unity in the momentum basis at the initial time and a further resolution of unity in the position basis at the final (echo) time, we obtain
\begin{align}
\chi&=\int d^fp^-\, d^fp^+\, d^fX\, \braket{ \mathbf{z}_0  |\mathbf{p}^-} \braket{ \mathbf{p}^-| U^\dagger_{\beta-\varepsilon}|\mathbf{X}} \braket{ \mathbf{X}| U_{\beta+\varepsilon}
|\mathbf{p}^+} \braket{ \mathbf{p}^+| \mathbf{z}_0 },
\label{eq:echoamplapp}
\end{align}
where
\begin{align}
\label{eq:gwp}
\braket{ \mathbf{p}|\mathbf{z}_0 } = (\hbar\pi)^{-f/4}e^{-\frac{1}{2\hbar}(\mathbf{p}-\mathbf{p}_0)^2- \frac{i}{\hbar}\mathbf{x}_0 \cdot (\mathbf{p}-\mathbf{p}_0)}
\end{align}
in suitable units of momentum and position.
In the next step we make use of the mixed-representation Van Vleck-Gutzwiller (VVG) propagator \cite{Gutzwiller1990}
\begin{align}
\braket{ \mathbf{X}| U |\mathbf{p} }=(2 \pi \hbar)^{-f/2} |\det(M_{11})|^{-1/2} e^{i\frac{S(\mathbf{p},\mathbf{X})}{\hbar}-i\frac{\pi}{2} \mu} ,
\label{eq:vvgprop}
\end{align}
where $\hbar$ is the effective Planck constant, and
\begin{align}
M = \begin{pmatrix}\displaystyle
   \frac{\partial \mathbf{X}}{\partial \mathbf{x}} &
   \displaystyle\frac{\partial \mathbf{X}}{\partial \mathbf{p}}
   \\[.4cm]
   \displaystyle
  \frac{\partial \mathbf{P}}{\partial \mathbf{x}} & 
  \displaystyle\frac{\partial \mathbf{{P}}}{\partial \mathbf{p}}  \label{eq:M definition}
    \end{pmatrix} 
\end{align} 
is the classical stability matrix for a trajectory from initial conditions $(\mathbf{x},\mathbf{p})$ to final conditions $(\mathbf{X},\mathbf{P})$. 
This trajectory is generated by the classical action $S(\mathbf{p},\mathbf{X})$, while $\mu$ is the Maslov index, an integer depending on the path that keeps track of the branch of the square roots in the first factor of the VVG propagator. 

Substituting \eqref{eq:gwp} and the VVG propagator \eqref{eq:vvgprop} into the echo amplitude \eqref{eq:echoamplapp} yields
\begin{align}
\chi=&\frac{1}{2^f(\pi\hbar)^{3f/2}}\int d^fp^-\, d^fp^+\, d^fX\, \left| \det(M_{11}^+) \det(M_{11}^-) \right|^{-1/2} \exp\left(-\frac{1}{2\hbar}[(\mathbf{p}^+-\mathbf{p}_0)^2+(\mathbf{p}^- -\mathbf{p}_0)^2]+ \frac{i}{\hbar}R \right),
\end{align}
where the overall phase
\begin{align}
R=\mathbf{x}_0 \cdot (\mathbf{p}^- -\mathbf{p}^+) + S(\mathbf{p}^+,\mathbf{X},\beta+\varepsilon)-S(\mathbf{p}^-,\mathbf{X},\beta-\varepsilon)
\end{align}
is antisymmetric in $(\mathbf{p}^+,\beta+\varepsilon) \leftrightarrow(\mathbf{p}^-,\beta-\varepsilon)$, and now we also make explicit the dependence of $S$ on the parameter $\beta$. 
Furthermore, we already anticipated that in the limit $\varepsilon\to 0$, the Maslov indices generally cancel.

To systematically expand around $\varepsilon=0$, we write $\mathbf{p}^\pm=\mathbf{p}\pm \tilde{\mathbf{p}}/2$, $\mathbf{x}_\pm=\mathbf{x}\pm \tilde{\mathbf{x}}/2$, giving
\begin{align}
R\approx -\mathbf{x}_0 \cdot \tilde{\mathbf{p}}+2\varepsilon \: \partial_{\beta} S(\mathbf{p},\mathbf{X},\beta)+\mathbf{x} \cdot \tilde{\mathbf{p}} ,
\end{align}
so that
\begin{align}
\chi\approx&\frac{1}{2^f(\pi\hbar)^{3f/2}}\int
d^fp\, d^fX \,d^f\tilde{p}\,
\left| \det(M_{11}^+) \det( M_{11}^-) \right|^{-1/2} \exp\left(-\frac{(\mathbf{p}-\mathbf{p}_0)^2}{\hbar}-\frac{\tilde{p}^2}{4\hbar}+ \frac{i}{\hbar}[(\mathbf{x}-\mathbf{x}_0)\cdot\tilde{\mathbf{p}}+2\varepsilon \: \partial_{\beta} S] \right)
\label{eq:intermediatestep1} 
\\
\approx
&\frac{1}{(\pi\hbar)^f}\int d^fp \,d^fX \, \left| \det(M_{11}^+)  \det(M_{11}^-) \right|^{-1/2}
\exp\left(-\frac{(\mathbf{x}-\mathbf{x}_0)^2+(\mathbf{p}-\mathbf{p}_0)^2}{\hbar}+ \frac{i}{\hbar}2\varepsilon \: \partial_{\beta} S\right) .
\label{eq:intermediatestep2} 
\end{align}
The second line corresponds to a saddle point approximation in leading order of $\hbar^{-1}$, in which the stability factors involving $M_{11}^\pm$ are evaluated at $\tilde{\mathbf{p}}=0$. We can then equate these stability factors from Eq.~\eqref{eq:M definition} as
\begin{align}
\lim_{\varepsilon\to 0}
\left| \det(M_{11}^+) \det(M_{11}^-) \right|^{-1/2}= \left| \det\left(\frac{\partial \mathbf{x}}{\partial\mathbf{X}}\right)\right|\end{align}
with the Jacobian of the change of variables from the final position $\mathbf{X}$ to the initial position $\mathbf{x}$.
This allows us to rewrite the echo amplitude
\begin{align}
\chi&\approx \frac{1}{(\pi\hbar)^f}\int d^f\!p \, d^f\!X \, \left| \det\left(\frac{\partial \mathbf{x}}{\partial \mathbf{X}}\right)\right| \exp\left(-\frac{(\mathbf z-\mathbf z_0)^2}{\hbar}+ \frac{i}{\hbar}2\varepsilon \: \partial_{\beta} S\right)=\frac{1}{(\pi\hbar)^f}\int d^{2f}\!\mathbf{z} \,\exp\left(-\frac{(\mathbf z-\mathbf z_0)^2}{\hbar}+ \frac{i}{\hbar}2\varepsilon \: \partial_{\beta} S\right)
\end{align}
in terms of the initial phase space points $\mathbf{z}=(\mathbf{x},\mathbf{p})$. Here, $\partial_{\beta} S$ signifies the partial derivative of the classical action. 

Finally, using Eq.~\eqref{eq:QFI1APP}, by differentiating the echo amplitude with respect to $\varepsilon$ and then setting $\varepsilon=0$, we arrive at the expression
\begin{align}
I_{{sc}}(\mathbf z_0)=\frac{4}{\hbar^2}
\left[\frac{1}{(\pi\hbar)^f} \int d^{2f}\!\mathbf{z}
 \; e^{-\frac{(\mathbf{z}-\mathbf{z}_0)^2}{\hbar}} \: (\partial_{\beta} S)^2 -
\left(\frac{1}{(\pi\hbar)^f} \int d^{2f}\!\mathbf{z}
\; e^{-\frac{(\mathbf{z}-\mathbf{z}_0)^2}{\hbar}} \: \partial_{\beta} S \right)^2\right]
\label{eq:mainVVG2}.
\end{align}
This can then be expressed as a variance
\begin{align}
I_{\mathrm{sc}}(\mathbf z_0)&=\frac{4}{\hbar^2}\mathrm{var}\left(\frac{\partial S}{\partial \beta}\right)
\label{eq:mainVVG11}
\end{align}
obtained by sampling initial phase space points $\mathbf{z}$ according to the distribution
\begin{equation}
W(\mathbf{x},\mathbf{p})=\frac{1}{(\pi\hbar)^f} d^{2f}\mathbf{z}
\exp\left[-\frac{(\mathbf{z}-\mathbf z_0)^2}{\hbar}\right],
\end{equation}
corresponding to a Gaussian centered at $\mathbf z_0$ with a covariance matrix $\Sigma=\frac{\hbar}{2}\openone$, where $\openone$ is the $2f$-dimensional identity matrix, that conforms to minimal uncertainty.

Revisiting the steps in this derivation for an arbitrary initial state $\ket{\psi_0}$, Eq.~\eqref{eq:intermediatestep1} is modified to read
\begin{align}
\chi\approx&\frac{1}{(2\pi\hbar)^{f}}\int
d^fp\, d^fX \,d^f\tilde{p}\,
\left| \det(M_{11}^+) \det( M_{11}^-) \right|^{-1/2} \braket{ \psi_0|\mathbf{p}-\tilde{\mathbf{p}}/2}\braket{ \mathbf{p}+\tilde{\mathbf{p}}/2|\psi_0} 
\exp\left( \frac{i}{\hbar}[\mathbf{x}\cdot\tilde{\mathbf{p}}+2\varepsilon \: \partial_{\beta} S] \right)
\label{eq:intermediatestep1general} 
\\
\approx
&\int d^fp \,d^fx \,
W(\mathbf{x},\mathbf{p})\,\exp\left( \frac{i}{\hbar}2\varepsilon \: \partial_{\beta} S\right) ,
\label{eq:intermediatestep2general} 
\end{align}
where the second line again corresponds to a saddle point approximation around $\tilde {\mathbf{p}}=0$, and
\begin{align}
W(\mathbf{x},\mathbf{p})&=
\frac{1}{(2\pi\hbar)^f}\int d^f\tilde{p}\,
\braket{\psi_0|\mathbf{p}-\tilde{\mathbf{p}}/2} \braket{\mathbf{p}+\tilde{\mathbf{p}}/2|\psi_0} 
\exp\left(\frac{i}{\hbar}\mathbf{x}\cdot\tilde{\mathbf{p}}\right) 
\nonumber
\\
&=
\frac{1}{(2\pi\hbar)^f}\int d^f\tilde{x}\,
\braket{\psi_0|\mathbf{x}+\tilde{\mathbf{x}} /2} \braket{\mathbf{x}-\tilde {\mathbf{x}}/2|\psi_0} 
\exp\left(\frac{i}{\hbar}\tilde{\mathbf{x}}\cdot{\mathbf{p}}\right) 
\end{align}
is now given by the Wigner function of the initial state. The semiclassical QFI remains of the form $I_{\mathrm{sc}}=\frac{4}{\hbar^2}\mathrm{var}\left(\frac{\partial S}{\partial \beta}\right)$, where the variance is now defined with respect to this general quasiprobability distribution \footnote{The implications for nonclassical initial states will be addressed elsewhere.}.

\section{Numerical Approach to QFI in the Quantum Kicked Top}
\subsection{Exact numerical approach}
We now describe in detail the numerical computational method to obtain the exact QFI in the quantum kicked top (KT). As specified in Eq.~(5) of the main text, the time evolution operator for the KT is $U_{\beta}(t\:T)=U_{\beta}(T)^t$ with the Floquet operator
\begin{equation}
U_{\beta}(T)
=\exp{(-ik\frac{J_z^2}{2J+1})}\exp{(-i\beta J_y)}
.
\end{equation}
For notational convenience, we set $T\equiv 1$ and denote $U_{\beta}(T)\equiv U_\beta$.
For a fixed $J$, the Hilbert space of dimension $2J+1$ is spanned by the eigenbasis $\ket{J,m}$, where $m=-J,-J+1,\ldots, J$, of the spin operator $J_z$. The elements of spin operators $J_z$ and $J_y=i(J_+^\dagger-J_+)/2$ are obtained from
\begin{align}
    (J_z)_{mn}&=m\:\delta_{m,n},
    \\
    (J_{+})_{mn}&=\sqrt{(J+m)(J-n)}\:\delta_{m-1, n}\,,
\end{align}
where $\delta_{m,n}$ is the Kronecker delta symbol.
Using the unitary matrix $V_y$ that diagonalizes the operator $J_y$ as
\begin{align}
    V_y^\dagger J_y V_y = J_z
    ,
\end{align}
we can furthermore rewrite
\begin{align}
    U_{\beta}=\exp{(-ik\frac{J_z^2}{2J+1})} V_y \exp{(-i\beta J_z)} V_y^\dagger ,
    \label{eq: U_0}
\end{align}
where the exponential terms can be easily calculated, as they only contain diagonal matrices $J_z$ and $J_z^2$.

To evaluate the parametric derivative of the Floquet operator with respect to $\beta$ we use $\partial_{\beta}U_{\beta}=-iU_{\beta}J_y$, and apply the product rule to obtain
\begin{align}
    \partial_{\beta}U_{\beta}(t)=\sum_{i=0}^{t-1} U_{\beta}^{i} (-i U_{\beta} J_y) U_{\beta}^{t-i-1}
    \label{eq: derivative of F}
    .
\end{align}
To further optimize the computation of \eqref{eq: derivative of F}, we decompose the time $t$ into a sequence of steps $2^u$, with integer $u$, and set $G_{\beta}(u)=U_{\beta}(2^{u})$. For each of these steps, Eq.~\eqref{eq: derivative of F} can then be implemented efficiently using the recursion relations
\begin{eqnarray}
    \partial_{\beta}G_{\beta}(u+1)&=&\partial_{\beta}G_{\beta}(u) \; G_{\beta}(u) + G_{\beta}(u) \; \partial_{\beta}G_{\beta}(u) ,
    \label{eq: derivative of F with trick}\\
    G_{\beta}(u+1)&=&G_{\beta}(u) \; G_{\beta}(u) ,
\end{eqnarray}
in that order, with initial conditions $\partial_{\beta}G_{\beta}(0)=-i U_{\beta} J_y$ and $G_{\beta}(0)=U_{\beta}$.

The QFI is then obtained by inserting these expressions into
\begin{align}
I_{\beta}(\ket{\mathbf z_0},t) = 4 \, (\, \braket{\mathbf z_0|\partial_{\beta} U_{\beta}^\dagger(t) \: \partial_{\beta} U_{\beta}(t)|\mathbf z_0}-|\braket{\mathbf z_0|U_{\beta}^\dagger(t)\: \partial_{\beta} U_{\beta}(t)|\mathbf z_0}|^2\, )
\label{eq:QFI}
,
\end{align}
where the matrix elements of the initial states $\ket{\mathbf{z}_{0}}=\ket{J,\theta,\phi}$, with $\theta=\arccos{(z)}$, are evaluated by direct implementation of the $\mathrm{SU}(2)$ spin-coherent states
\begin{align}
\label{eq:spin_coherent_state1}
\ket{J,\theta,\phi}=\sum_{m=-J}^{J}\sqrt{
\begin{pmatrix}
2J\\
J-m
\end{pmatrix}
}[e^{i\phi}\sin{(\frac{\theta}{2})}]^{J-m}\cos{(\frac{\theta}{2})^{J+m}}\ket{J,m}
,
\end{align}
also given in Eq.~(9) of the main text.

\subsection{Semiclassical approach} 
To evaluate the semiclassical QFI \eqref{eq:mainVVG11} for the KT, we need to determine the parametric derivative of the action of classical trajectories, and obtain their variance from the phase space integral \eqref{eq:mainVVG2}. 

\subsection{Parametric derivative of the action}
The classical evolution consists of linear rotations by an angle $\beta$ about the $y$-axis, and a nonlinear torsion by an angle $k\,z$ about the $z$-axis \cite{haake1987classical}.
Over a single time step of the classical time evolution, the corresponding action generating this dynamics in the phase space $(\phi,z)$ is given by \cite{Braun1996}
\begin{equation}
S^{(1)}(z_t,z_{t+1})= z_t \arccos{\frac{z_t \cos{\beta}-z_{t+1}}{\sin{\beta}\sqrt{1-z_t^2}}} - z_{t+1} \arccos{\frac{z_t-z_{t+1} \cos{\beta}}{\sin{\beta}\sqrt{1-z_{t+1}^2}}} + \arccos{\frac{z_t\, z_{t+1}-\cos{\beta}}{\sqrt{(1-z_t^2)(1-z_{t+1}^2)}}} - \frac{k}{2}z_{t+1}^2,
\end{equation}
where the first three terms generate the linear rotation and the last term generates the nonlinear torsion.
From this, we obtain the parametric derivative with respect to $\beta$ as
\begin{align}
\partial_{\beta} S^{(1)}(z_t,z_{t+1}) = -\sqrt{1-z_t^2}\sin{\phi_t}=-y_t\:,
\end{align}
where $y_t$ is the Cartesian coordinate of the unit vector in the direction of the classical spin. This proves Eq.~(6) of the main text. Over $t$ time steps, the action of a trajectory can be obtained according to the composition law
\begin{align}
S(t) = \sum_{t'=0}^{t-1} S^{(1)}(z_{t'},z_{t'+1}) 
\label{eq:action_summation11}
.
\end{align}
We therefore obtain Eq.~(7) of the main text,
\begin{align}
\partial_{\beta} S(t) = \sum_{t'=0}^{t-1} \partial_{\beta} S^{(1)}(z_{t'},z_{t'+1}) = -\sum_{t'=0}^{t-1} y_{t'}
\label{eq:action_summation1}
.
\end{align}

\subsection{Phase space discretization}
\begin{figure*}[t] 
    \centering
    \includegraphics[width=0.4\textwidth]{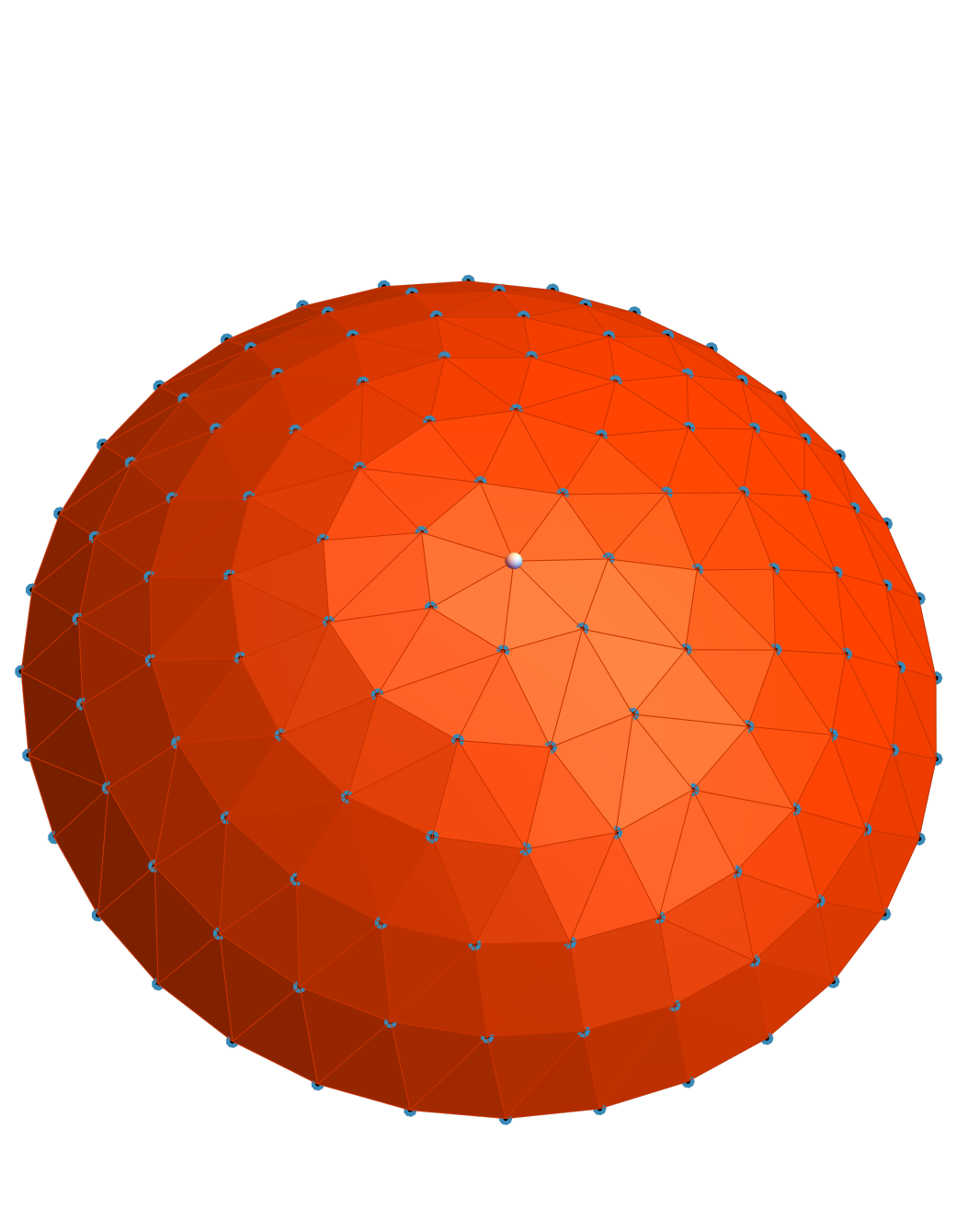}
    \caption{Illustration of the construction of grid points utilized to evaluate the semiclassial QFI $I_{sc}$ around a central point (white dot). Denoting the arc distance between neighboring points as $d_0$, and assuming that the radius of the shown cap--- measured in arc length--- equals the width $\sigma$ of the evaluated Gaussian distribution, the figure corresponds to a phase space resolution parameter $r=\sigma/d_0=6$. In the numerical implementation the cap extends to an arc length $R_\mathrm{eff}=5\sigma$, while $r=50$, unless stated otherwise, resulting in approximately $\approx 1.96\times 10^5$ grid points.
    }
    \label{fig:Gaussian distribution}
\end{figure*}

For the numerical evaluation of the phase space integrals in Eq.~\eqref{eq:mainVVG2}, we have to evaluate a Gaussian phase space distribution corresponding to the spin-coherent initial state \eqref{eq:spin_coherent_state1}. As shown in Fig.~\ref{fig:Gaussian distribution}, this integral is most conveniently discretized in Cartesian coordinates with points located on the cap of the unit sphere surrounding the polar-coordinate point $(\phi,z)$.
The discretization grid is determined by three distances:
(i) The width of the Gaussian wave packet 
\begin{align}
    \sigma=1/\sqrt{2J+1}\,,
\label{eq:sigma}
\end{align}
(ii) the arc distance $d_0$ between neighboring grid points, and (iii) the cutoff range $R_\mathrm{eff}\equiv 5\sigma$ of the integration region.
To enable the comparison of data at different values of $J$, we introduce the phase space resolution parameter
\begin{align}
    r=\sigma/d_0 \,,
\end{align}
which specifies the number of grid points within range $\sigma$ in one direction, and will be kept as an integer value.
Then, the number of grid points within arc distance $r$ around the main point can be obtained from
\begin{align}
    N(r)=1+\sum_{i=1}^{r} \lceil 2\pi i \rceil ,
\end{align}
where $\lceil \cdot \rceil$ indicates the ceiling function. Generally, $N(r)>\pi r^2$, but as $r$ increases, the two values converge, since at infinity we have
\begin{align}
    \lim_{r\to \infty} N(r) / r^2=\pi.
\end{align}
In Fig.~3 of the main text we observe a practical convergence of the numerical for $r\gtrsim 20$. The numerical results in the main text and the present supplemental material have been obtained for $r=50$, corresponding to $\sim (R_\mathrm{eff}/\sigma)^2\pi r^2\approx 1.96\times 10^5$ grid points.

With these specifications, the initial trajectories are then evaluated using Gaussian weights
\begin{align}
   w_i = \frac{1}{\mathcal{N}} \exp{\left( -\frac{(2J+1)d_{i0}^2}{2} \right)} ,
\end{align}
where $d_{i0}$ is the arc distance between the grid point $i$ and the center of the cap, and the normalization factor is obtained from
\begin{align}
    \mathcal{N}=\sum_i \exp{\left( -\frac{(2J+1)d_{i0}^2}{2} \right)}.
\end{align}

Collecting all results, the semiclassical QFI in the KT then takes the explicit form
\begin{align}
I_{sc}(\mathbf z_0, t) = \left(2J+1\right)^2 \left( \left<\left(\sum_{t'=0}^{t-1} y_{t'}\right)^2\right> - \left<\sum_{t'=0}^{t-1} y_{t'}\right>^2 \right) ,
\end{align}
where the numerical averages are given by
\begin{align}
    \left< A \right> := \sum_i w_i A_i \,.
\end{align}

\section{Detailed Comparison of $I_\text{sc}$ and $I$ Using Additional Data}
\begin{figure*}[t] 
    \centering
    \hspace{-0.16cm} 
    \subfigure{
        \includegraphics[width=0.25\textwidth]{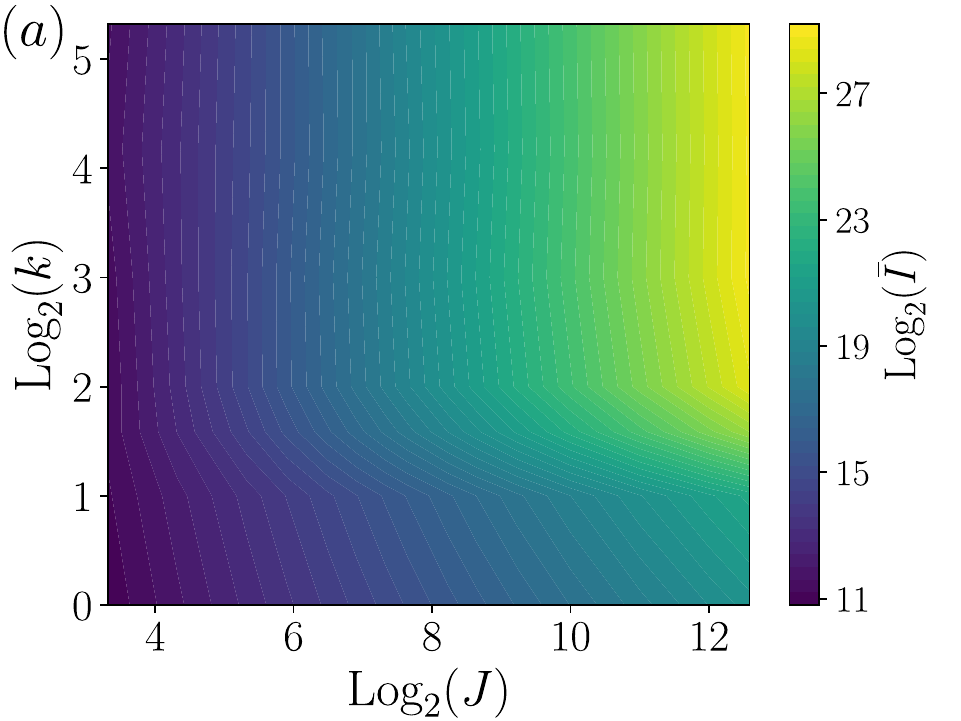}
        }\subfigure{
    \includegraphics[width=0.25\textwidth]{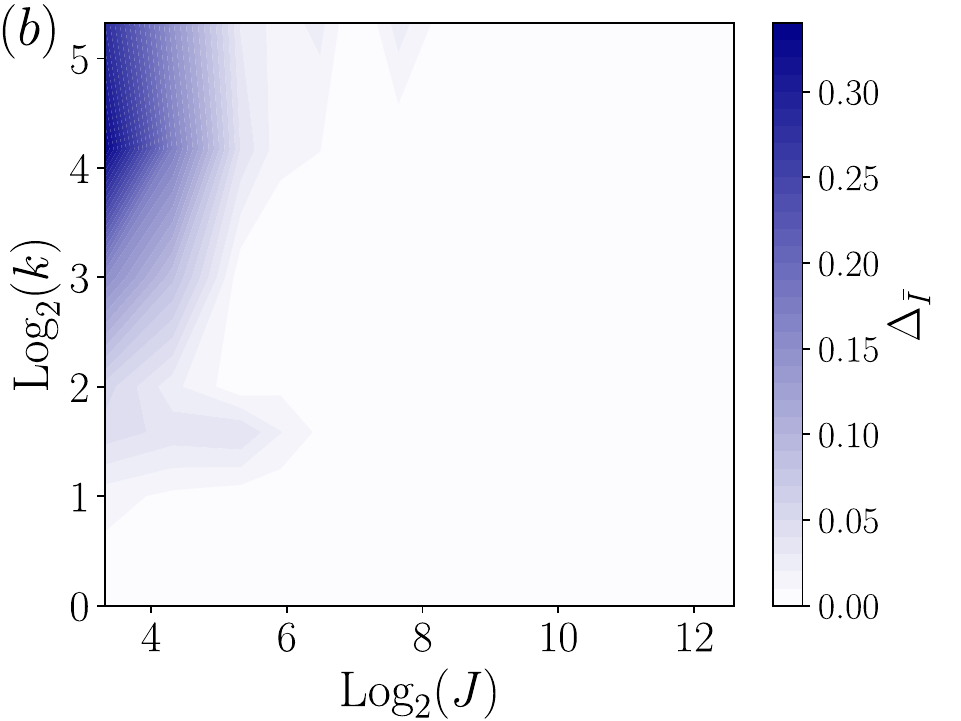}
    }\subfigure{
    \includegraphics[width=0.25\textwidth]{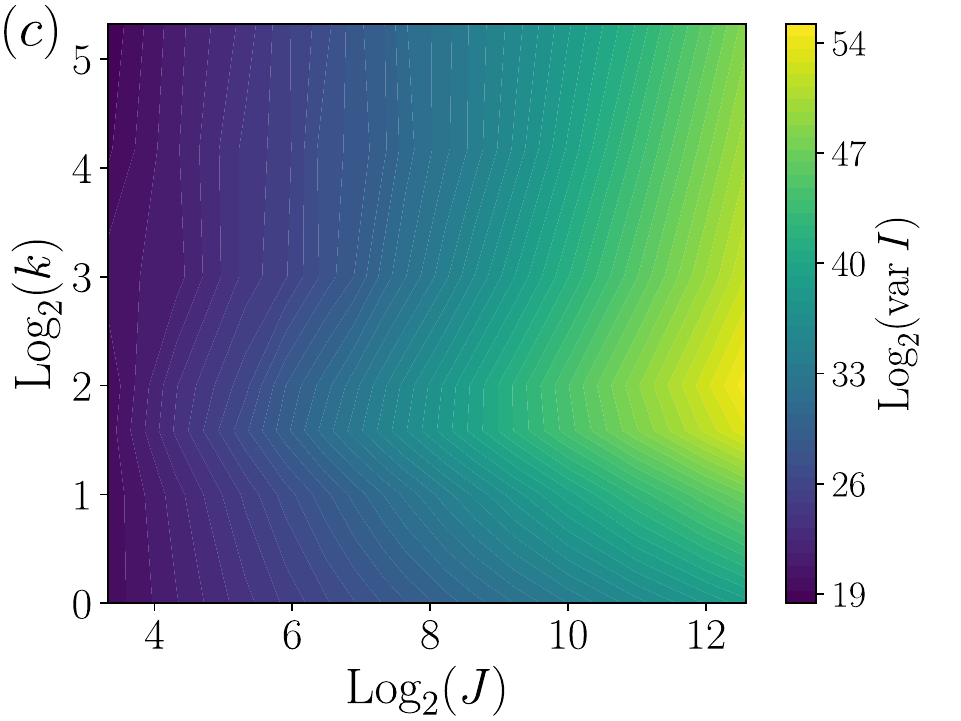}
    }\subfigure{
       \includegraphics[width=0.25\textwidth]{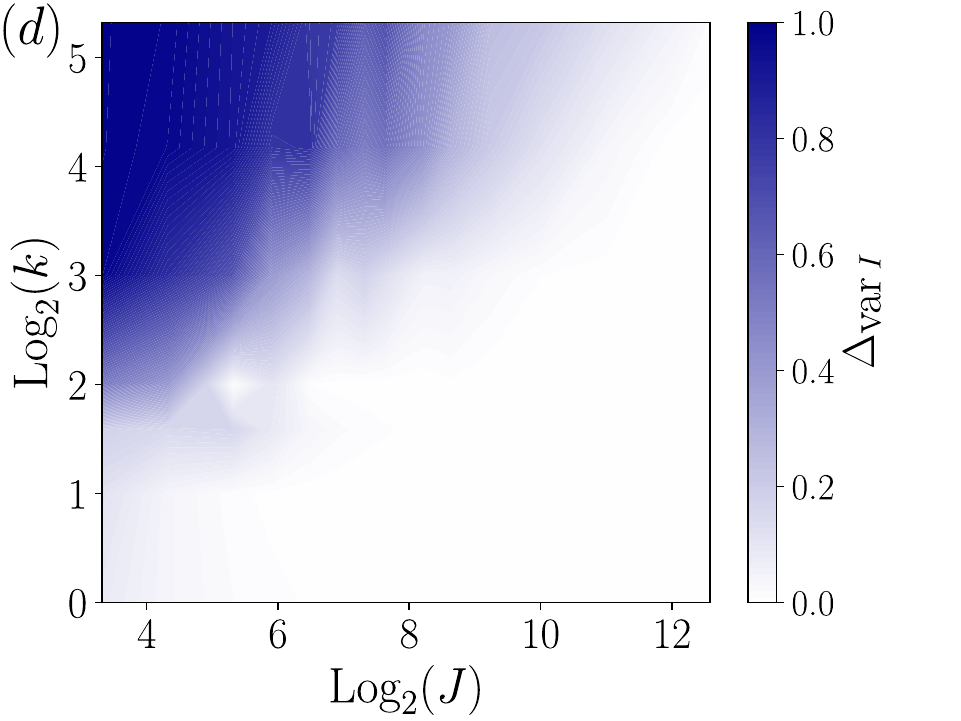}
    }
    \caption{Comparison of the exact and semiclassical results for the QFI in terms of the phase space average $\bar I$ and the phase space variance $\mathrm{var}\, I$, in analogy to Fig.~2 of the main text, but as a function of $J$ and $k$ for fixed time $t=12$. 
    Excellent agreement is obtained in the semiclassical regime of large $J$, and this agreement is more readily attained for small and moderate values of $k$.}
    \label{fig:ave_var_comp_fixed_t}
\end{figure*}

\begin{figure*}[t] 
    \centering
    \hspace{-0.16cm} 
    \subfigure{
        \includegraphics[width=0.25\textwidth]{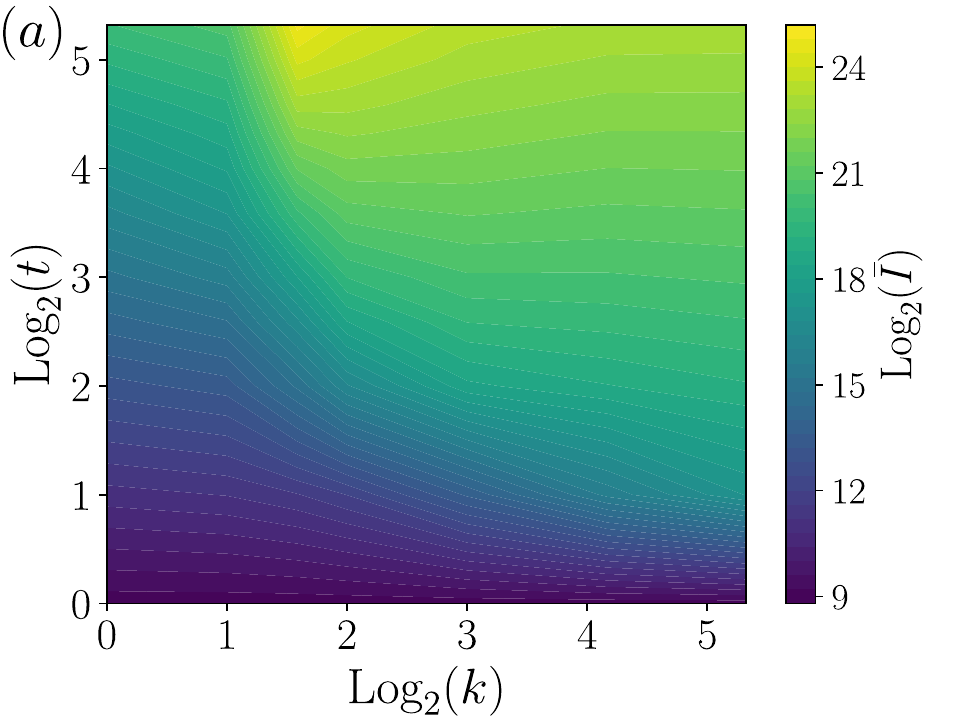}
    }\subfigure{   \includegraphics[width=0.25\textwidth]{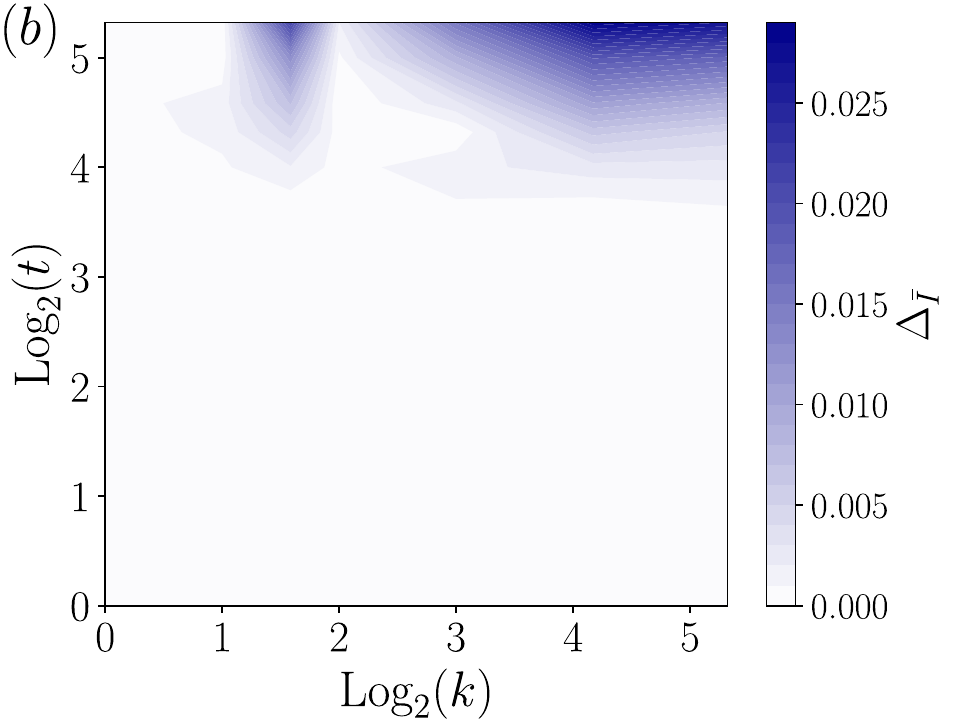}
    }\subfigure{   \includegraphics[width=0.25\textwidth]{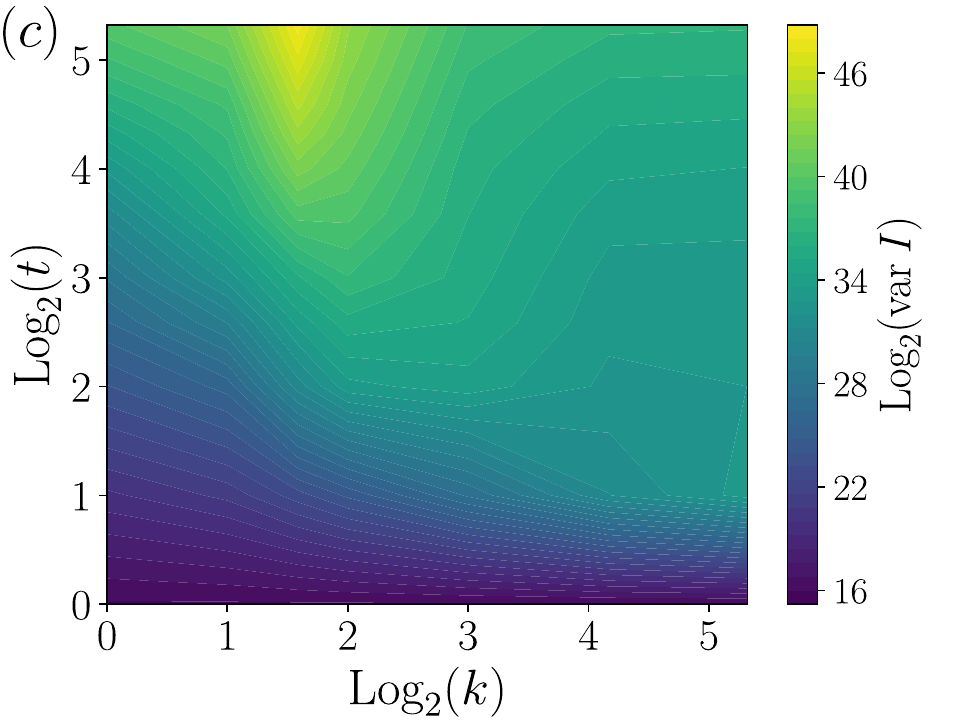}
    }\subfigure{     \includegraphics[width=0.25\textwidth]{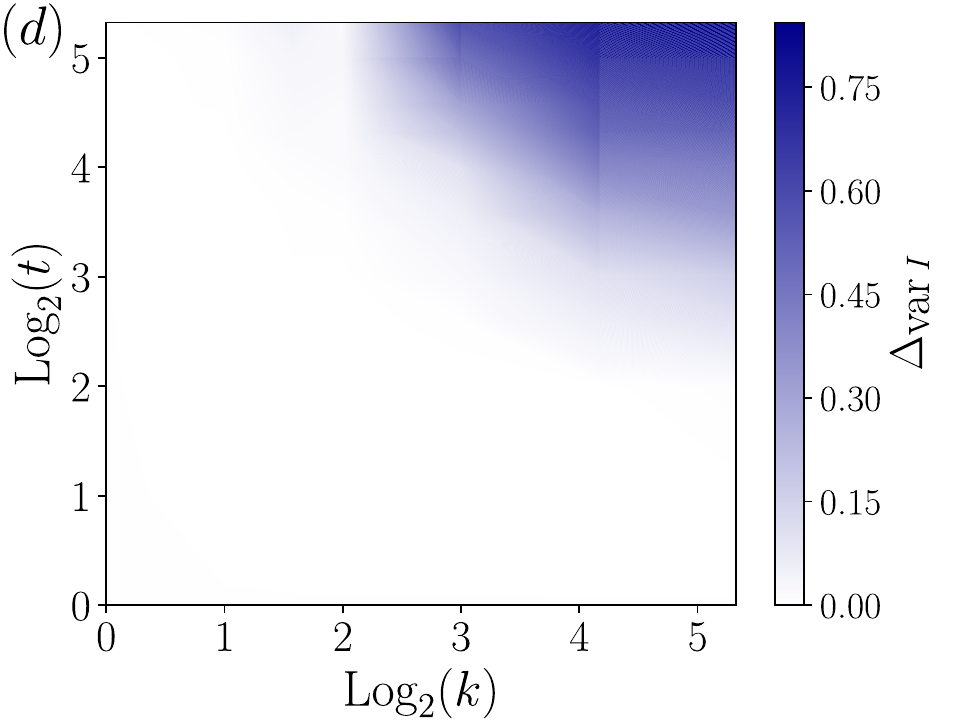}
    }
    \caption{Comparison of approaches in analogy to Fig.~2 of the main text and Fig.~\ref{fig:ave_var_comp_fixed_t} in this supplemental material, but as a function of  $k$ and $t$ with fixed $J=400$.
    Excellent agreement is observed up to the limitations set by the Heisenberg time (in the direction of increasing time) and the  numerical resolution of the chaotic phase space (in  the direction of increasing $k$.)
    }
    \label{fig:ave_var_comp_fixed_J}
\end{figure*}

Here, we provide additional data for the comparison of the semiclassical method with the exact QFI. We again use the quantities $\bar I$ and $\mathrm{var}\, I$ for the entire phase space, and quantify the comparison by the corresponding deviation $\Delta_X \equiv |\bar X-\bar X_{\mathrm{sc}}|/(\bar X+\bar X_{\mathrm{sc}})$. These data are presented as functions of the variables $J$, $t$, and $k$ while keeping $\beta=1.5$ fixed. We apply the following ranges for the variables: $J\in\{10,20,40,60,90,120,160,200,300,400,600,1024,2048,4096,6144\}$, $t\in\{1,2,4,6,8,12,16,20,24,32,40\}$, and $k\in\{1,2,3,4,8,18,40\}$. The initial coherent state is parameterized in terms of phase space variables $\phi\in(0,2\pi)$ and $z\in(-1,1)$, which are discretized using 80 and 55 equidistant points, respectively, in the $\phi$ and $z$ directions. The case involving a fixed $k=4$ has already been addressed in the main text. Therefore, we present here the case of fixed $t=12$, Fig.~\ref{fig:ave_var_comp_fixed_t}, as well as fixed $J=400$, Fig.~\ref{fig:ave_var_comp_fixed_J}.

Within the parameter range where the semiclassical method can be applied efficiently, the method again proves accurate. 
At fixed $J$ and $t$, a computational limit is reached for large $k$, characterized by a highly chaotic regime with a very large Lyapunov exponent. The computational effort to resolve the features of the classical dynamics then increases exponentially, leading to inaccurate results in the quantum regime. However, for large but fixed values of $k$ and $t$, the method recovers its accuracy as $J$ is increased, so that smaller regions of phase space need to be evaluated. This underlines the semiclassical nature of the approach.

\begin{figure}[t] 
    \centering
    \subfigure{
        \includegraphics[width=0.28\linewidth]{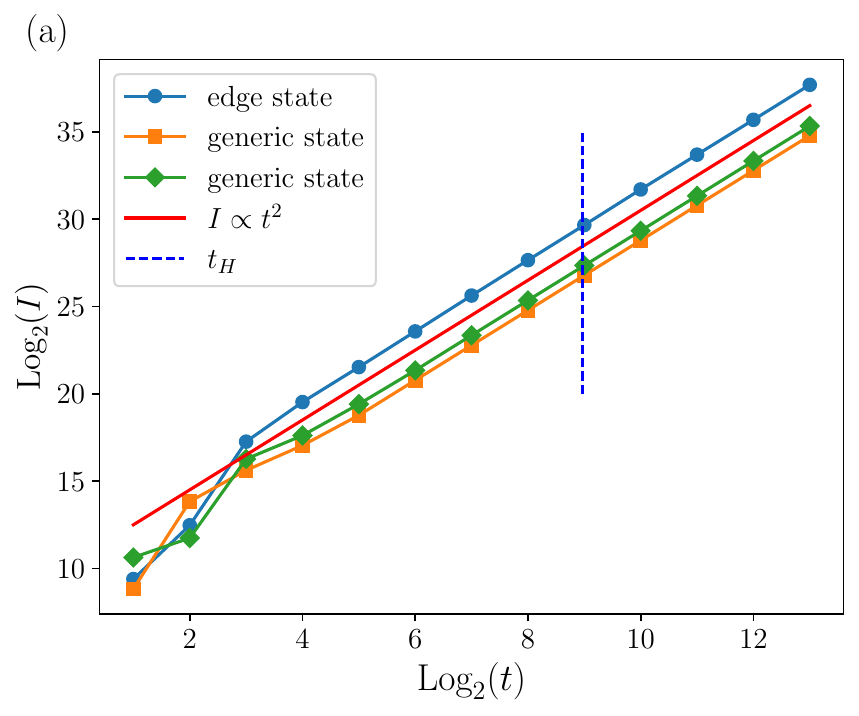}
    }\hfill
    \subfigure{
        \includegraphics[width=0.28\linewidth]{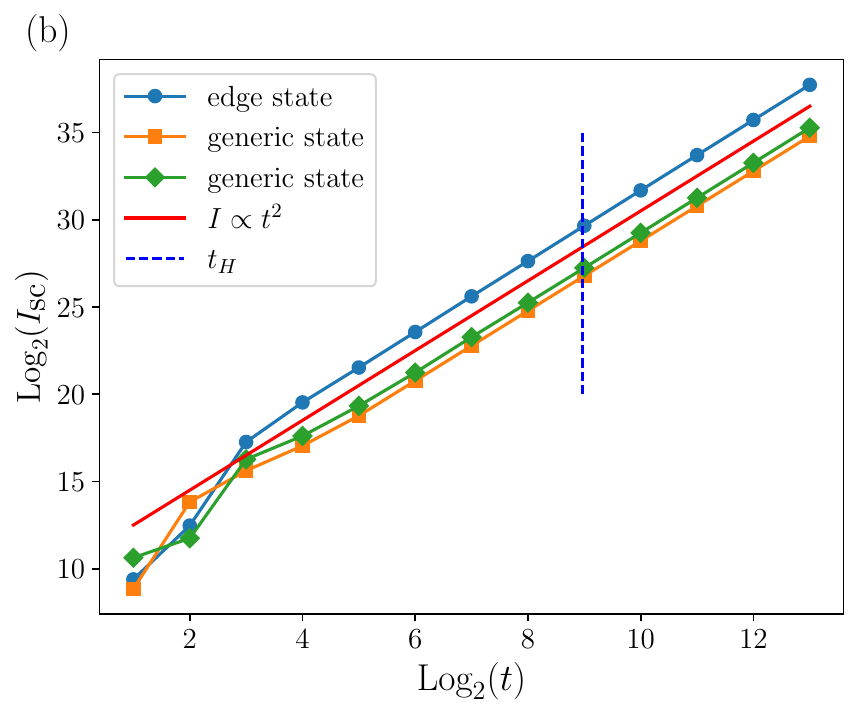}
   }\hfill
    \subfigure{
        \includegraphics[width=0.28\linewidth]{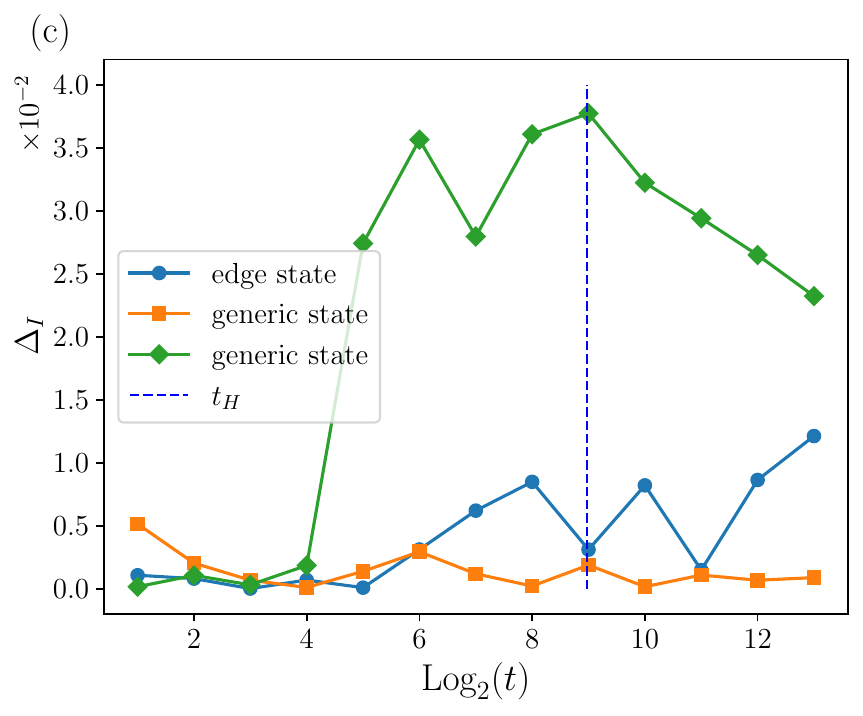}
    }
    \caption{Comparison of the exact QFI (a) with the semiclassical QFI (b) for the QKT for $J=250$, $\beta=1.5$, and $k=2$ for three representative initial states. The red line indicates scaling $I\propto t^2$ and the blue dashed line is the Heisenberg time. (c) Relative difference $\Delta_I$ as defined in Eq.~(10) of the main text.}
    \label{fig:1a}
\end{figure}

For further illustration of the accuracy of the method for specific states, we give in Fig.~\ref{fig:1a} data showing the QFI for three different initial states (one state localized at the edge of a stability island where the QFI is very large, and  two states localized at generic locations) for $\beta=1.5$, $k=2$, and $J=250$, covering times beyond the Heisenberg time, with $\log_2(t_H)\approx 9$. 
The semiclassical approximation works well, with relative errors less than $\approx 1\%$ for the edge state and one of the generic initial states and less than $\approx 4\%$ for the other generic state.

\section{Application to other model systems}
We here illustrate the application of the semiclassical method to two additional model systems, the quantum kicked rotor and the H{\'e}non Heiles system.
\subsection{The quantum kicked rotor}
The quantum kicked rotor 
\cite{IZRAILEV1990299} describes the free rotation of a degree of freedom on a ring $0\leq x<2\pi$ with a periodic boundary condition, supplemented by stroboscopic forces $\sim k\sin(x)$ acting at intervals $T$, where $k$ is the nonlinearity parameter. 
Classically, the system affords a description with a Hamiltonian $H=p^2/2m+k\sum_n \delta(t/T-n)\cos(x)$ that splits into a kinetic energy term and a stroboscopic potential. Setting $m=1$, the resulting classical evolution over one kicking period $T\equiv 1$ is the standard map
\begin{equation}
x_{t+1}=x_t +p_t+ k\sin x_t \pmod {2\pi},\qquad p_{t+1}=p_t+ k\sin x_t\pmod {2\pi},
\label{eq:standardmap}
\end{equation}
where we confined $0\leq p<2\pi$ in the standard way so that the phase space $\mathbf{z}=(x,p)$ forms a torus. 
Quantum-mechanically, the corresponding
time evolution operator over one kicking period is given compactly by \cite{tworzydlo2003dynamical}
\begin{equation}
    [U_k(1)]_{mn}=\frac{1}{\sqrt{iM}}\exp\left[\frac{i\pi}{M}(m-n)^2-ik\frac{M}{2\pi}\cos(2\pi n/M)\right],
\end{equation}
where $n,m=1,2,\ldots M$ with Hilbert space dimension $M$ and effective reduced Planck constant $\hbar_\mathrm{eff}=2\pi/M$. The dynamics for later times follows again from  $U_k(t)=U_k^t(1)$.

\begin{figure}
    \centering
    \includegraphics[width=\linewidth]{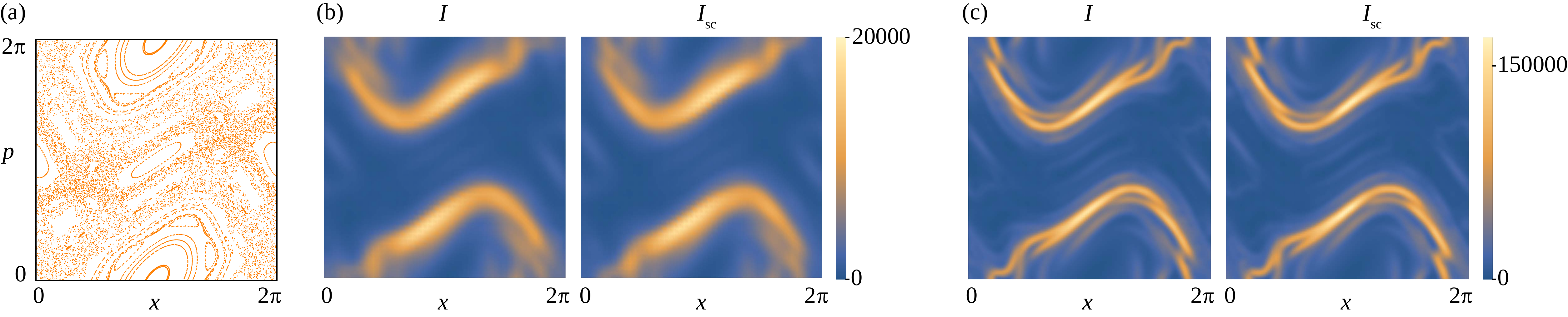}
    \caption{(a) Phase space portrait of the standard map \eqref{eq:standardmap} with $k=1.3$\:. (b) Phase-space resolved exact QFI (left) and semiclassical QFI (right) for the corresponding kicked rotor with $t=16$ and $M=100$.  (c)  Same as in (b), but for  $M=400$.}
    \label{fig:kr}
\end{figure}

In Fig.~\ref{fig:kr}, we compare the QFI   for the estimation parameter $k$, obtained quantum mechanically
with periodic coherent initial states, with the result from our semiclassical method, based on the action derivative
$\partial_{k} S(t) = \sum_{t'=0}^{t-1} \partial_{k} S^{(1)}(x_{t'},x_{t'+1})$ with $\partial_k S^{(1)}(x_{t},x_{t+1})=-\cos x_t$. 
We again find excellent quantitative and qualitative agreement. Furthermore, regions of large QFI correlate with the classical phase space structures in analogy to the QKT.

\begin{figure}[t]
    \centering
    \includegraphics[width=0.9\linewidth]{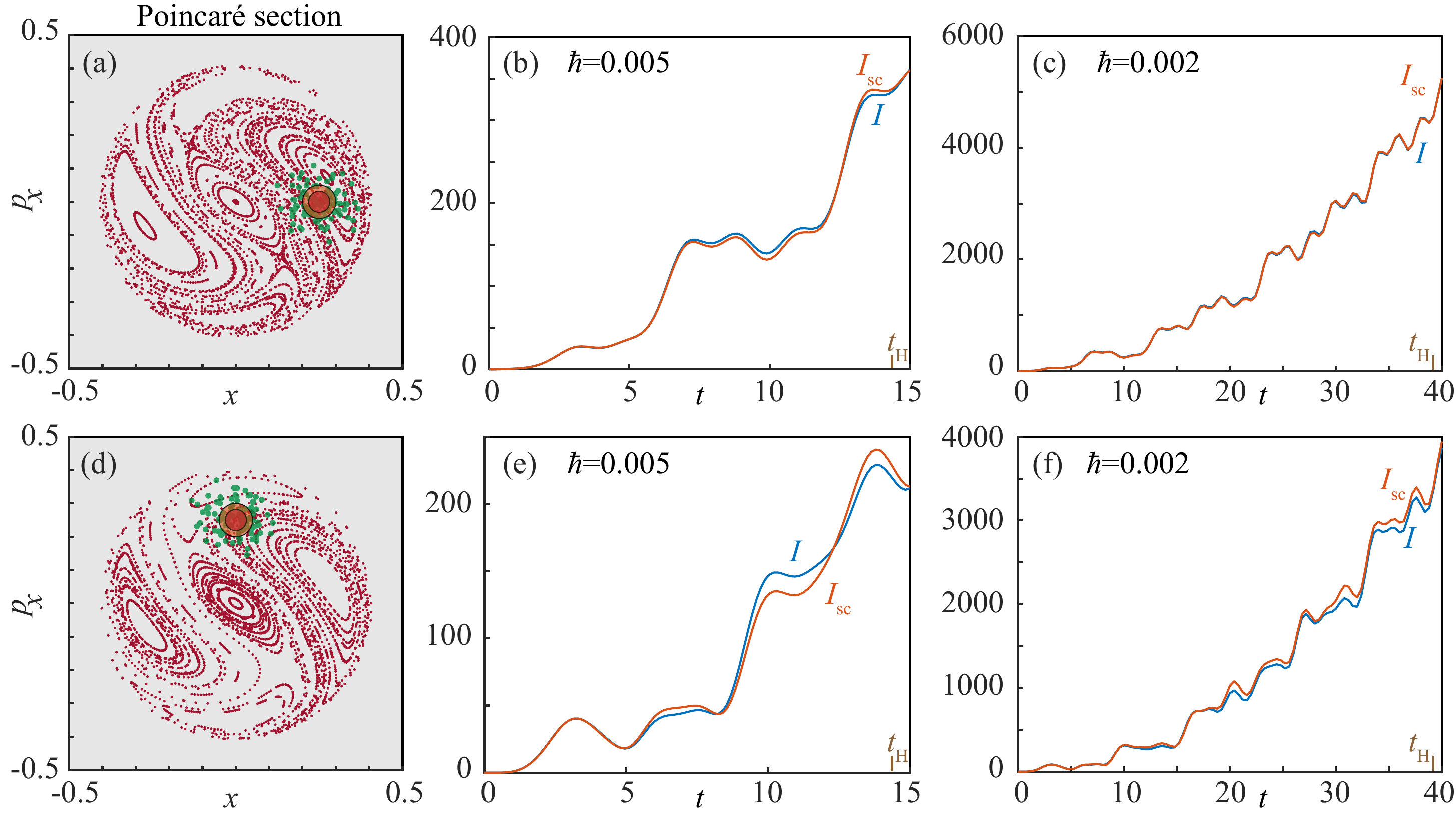}
    \caption{
    (a,d) Poincare surface of section  $y=0$, $p_y>0$ of the H{\'e}non-Heiles system \eqref{eq:HH} with $\lambda=1$ and 
    energy 
    $E=1/12$, indicating initial conditions for which the QFI is computed in panels (b,c,e,f). (b,c) Exact QFI $I$ (blue) and semiclassical QFI $I_{sc}$ (red) as a function of time for an initial coherent state located at the center of the disks in (a). (e,f) Analogously for the initial coherent state located at the center of the disks in (d).    
    In (b,e), $\hbar=1/200$, while in (c,f), $\hbar=1/500$. In (a,d), the semiclassical support of the initial states is indicated by disks of radius $\sqrt{\hbar/2}$, as well as the scatter of 100 random points obtained from the corresponding Gaussian distribution for the case $\hbar=1/200$. In (b,c,e,f), the semiclassical results are obtained from 50\,000 trajectories, while the quantum results are obtained in a basis of 6400 harmonic-oscillator states. The indicated Heisenberg time $t_H=\hbar/\Delta$ is estimated from the mean spacing $\Delta$ of 10 energy levels around $E=1/12$.}
    \label{fig:hh}
\end{figure}

\subsection{H{\'e}non-Heiles system}
The H{\'e}non-Heiles system \cite{brack2018semiclassical} is an autonomous quantum-chaotic system with two degrees of freedom and Hamiltonian
\begin{equation}
    H=\frac{1}{2}(p_x^2+p_y^2+x^2+y^2)+\lambda\left(x^2y-\frac{1}{3}y^3\right).
    \label{eq:HH}
\end{equation}
We set $\lambda=1$ and the energy $E=1/12$, where the model displays a mixed phase space as shown via the Poincar{\'e} surface of sections in Fig.~\ref{fig:hh}, and compare the exact and semiclassical QFI for  $\hbar = 1/200$ and $\hbar = 1/500$ and two initial conditions, given by coherent states $\psi_0$ centered at (top) $x=1/4$, $p_x=0$, $y=0$, and (bottom) 
$x=0$, $p_x=1/4$, $y=0$,
with the initial $p_y>0$ in both cases determined by the energy. 
The semiclassical method reproduces the quantum result in detail and becomes highly accurate as $\hbar$ decreases.
In this regime, the quantum calculations rapidly become prohibitively expensive, as is typical for models with multiple degrees of freedom.

These results are obtained by the following methods. For the quantum model, we apply canonical quantization with the given values of $\hbar$ and operate in the eigenbasis of the two-dimensional harmonic oscillator corresponding to $\lambda=0$, truncated to 6400 basis functions, having confirmed that the results are converged.
The QFI $I=4(\langle\psi_0|\hat L^2|\psi_0\rangle-\langle\psi_0|\hat L|\psi_0\rangle^2)$ for estimation parameter $\lambda$ 
is then obtained from the generator 
\begin{align}
\hat  L=\sum_{n,m(n\neq m)}\frac{e^{i(E_n-E_m)t/\hbar}-1}{i(E_n-E_m)}|n \rangle \langle n|\hat H'|m\rangle \langle m|+\sum_n
\frac{t}{\hbar}|n \rangle \langle n|\hat H'|n\rangle \langle n|
\end{align}
where $|n \rangle$ are the numerically obtained energy eigenstates with energy $E_n$ and $\hat H'=\left(\hat x^2\hat y-\frac{1}{3}\hat y^3\right)$.
Semiclassically, the initial states are represented by Gaussian distributions with variance $\mathrm{var}\, x=\mathrm{var}\,p_x=\mathrm{var}\,y=\mathrm{var}\,p_y=\frac{\hbar}{2}$ and a cutoff $|\mathbf{z}-\mathbf{z}_0|^2<9$, 
while the action derivative
$
\partial_\lambda S(t)=-\int_0^t H'(\mathbf{z}(t'))\,dt'
$ is obtained by numerical integration.

% \bibliography{refs.bib}

% \end{document}

% for arXiv>

\end{document}